\documentstyle[12pt]{article}

\makeatletter
\@addtoreset{equation}{section}
\makeatother

\def\theequation{\thesection.\arabic{equation}}
\newcommand{\be}{\begin{equation}}
\newcommand{\ee}{\end{equation}}
\newcommand{\bee}{\begin{eqnarray}}
\newcommand{\beee}{\begin{array}}
\newcommand{\eee}{\end{eqnarray}}
\newcommand{\eeee}{\end{array}}

\newcommand{\gn}{\nu}
\newcommand{\gm}{\mu}

\newcommand{\gx}{\xi}
\newcommand{\ga}{\alpha}
\newcommand{\gb}{\beta}
\newcommand{\kk}{\xi}
\newcommand{\gga}{\gamma}

\newcommand{\M}{{\cal M}}
\newcommand{\ls}{\!\!\!\!\!\!}

\newcommand{\gd}{\delta}

\newcommand{\gvep}{\varepsilon}
\newcommand{\gs}{\sigma}
\newcommand{\bz}{\bar z}
\newcommand{\go}{\omega}

\newcommand{\nn}{\nonumber}
\newcommand{\half}{\frac{1}{2}}
\newcommand{\ptl}{\partial}

\newcommand{\p}{\partial}

\newcommand{\E}{{\cal E}}

\newcommand{\V}{{\cal V}}

\newcommand{\D}{{\cal D}}

\newcommand{\C}{{\cal C}}

\newcommand{\f}{\frac}
\newcommand{\B}{{\cal B}}

\begin{document}

\begin{center}
{\large\bf Higher Rank Conformal Fields in the ${\bf Sp(2M)}$
Symmetric\\ Generalized Space-Time}
\vglue 0.6  true cm
\vskip1cm

O.A. Gelfond$^1$ and M.A.~Vasiliev$^2$
\vglue 0.3  true cm

${}^1$Institute of System Research of Russian Academy of Sciences,\\
Nakhimovsky prospect 36-1,
117218,
Moscow, Russia
%\vskip1cm
\vglue 0.3  true cm

%\medskip
${}^2$I.E.Tamm Department of Theoretical Physics, Lebedev Physical Institute,\\
Leninsky prospect 53, 119991, Moscow, Russia
%\medskip
%\vskip2cm
\vskip1.5cm
\end{center}

\begin{abstract}
We study
various  $Sp(2M)$ invariant field equations corresponding to rank $r$
tensor products of the Fock (singleton) representation of $Sp(2M)$.
These equations are shown to
describe localization on  ``branes'' of different dimensions
embedded into the generalized space-time $\M_M$ with matrix (i.e.,
``central charge'') coordinates.
The case
of bilinear tensor product is considered in detail.
The conserved currents built from bilinears
of rank 1 fields in $\M_M$ are shown to satisfy the field equations of the
rank 2 fields in $\M_M$. Also,
the rank 2 fields in $\M_M$ are shown to be
equivalent to the rank 1 fields in $\M_{2M}$.
\end{abstract}

\section{Introduction}

The idea that the set of $4d$ massless fields of all spins
should admit some manifestly $Sp(8)$ invariant formulation
in the ten-dimensional space-time $\M_4$
with symmetric matrix coordinates $X^{\ga\gb}=X^{\gb\ga}$
$(\ga,\gb = 1\ldots 4$ are $4d$ Majorana spinor indices)
was originally put forward by Fronsdal \cite{Fr1}.
In \cite{BL,BLS} twistor world line particle models in $\M_4$
were studied which upon quantization give rise to a
wave function equivalent to the $sp(8)$ singleton identified in \cite{Fr1}
with the set of all $4d$ massless representations of $o(3,2)$.
 In \cite{BHS} it was shown that the set of free equations of motion for
massless fields of all spins in $4d$ Minkowski
space-time indeed exhibits
$Sp(8)$ symmetry, which acts locally at the infinitesimal level
and admits equivalent description  in $\M_4$.
In \cite{BLS,BHS} the proposed constructions were extended
to the generic case of $Sp(2M)$ and it was argued that higher values
of $M$ correspond to higher spin theories in higher dimensions.

The problem of formulating a  consistent nonlinear higher spin gauge
theory is of great interest in the context of revealing a most
symmetric phase of a theory of fundamental interactions. Recently
it was conjectured that higher spin gauge theories
appear as AdS/CFT duals \cite{JM,GKP,Wit1}
of the boundary conformal theories in
the limit $g^2N\to 0$ \cite{Sun,Wit,SS,KP}. To check this conjecture
it is necessary to develop a nonlinear theory of massless higher spins
in the full generality, which is a nontrivial problem solved by now
only for the $3d$ and $4d$ higher spin theories at the level of
equations of motion \cite{PV,4d}
\footnote{For more references on higher spin theories we
refer the reader to \cite{Gol,SS}. Some  higher spin cubic
interactions in $5d$ theories were constructed recently in
\cite{5d}. { In \cite{add}  nonlinear equations of motion in
the  space-time of any dimension were obtained.}}.
 To this end it is important to work out
a most efficient formalism
that utilizes symmetries of the model as much as possible.
Once conformal higher spin theories are shown to
exhibit higher symplectic symmetries, an important step is to
develop explicitly $Sp(2M)$ invariant formulation. This is achieved
by virtue of introducing the generalized space-time  $\M_M$, which is
a minimal space where $Sp(2M)$ acts geometrically as
 Fronsdal showed in \cite{Fr1} where $\M_M$ was realized
as the space of $M$-forms of minimal rank.
Equivalently $\M_M$ can be defined \cite{Mar} as the coset space
$Sp(2M)/P$, where $P$ is the parabolic subgroup of $Sp(2M)$ generated
by the generalized Lorentz transformations, dilatations and special
conformal transformations (see below). $\M_M$ defined this way
is analogous to the compactified Minkowski space in the case of
usual conformal group $SO(d,2)$. Usual Minkowski space is the big
cell of the compactified Minkowski space. Analogously,
in this paper we analyze
the problem in the big cell of $\M_M$, which is $R^{\f{M(M+1)}{2}}$
(using the same notation $\M_M$ for the big cell throughout this paper).
The infinite towers of massless fields of all spins
in four dimensions were shown in \cite{BHS} to be
described by one scalar and one spinor
(equivalently, one scalar superfield) in $\M_4$.

The $Sp(2M)$  generalized conformal symmetry transformations
in $\M_M$ are realized by the vector fields (see, e.g., \cite{BHS})
\bee \nn
 P_{\ga\gb} =-i \f{\p}{\p X^{\ga\gb}}\,,\qquad
L_\ga{}^\gb = 2i
 X^{\gb\gga} \f{\p}{\p X^{\ga \gga}}\,,\qquad
K^{\ga\gb}=  -i X^{\ga\gga}
X^{\gb{\eta}}\f{\p}{\p X
^{\gga{\eta}}}\,.
\eee
The (nonzero) $sp(2M)$ commutation relations are
\bee \nn
[L_\ga{}^\gb \,, L_\gga{}^\gd] = i\left( \gd^\gd_\ga L_\gga{}^\gb -
\gd^\gb_\gga L_\ga{}^\gd \right )\,,
\eee
\bee \nn
[L_\ga{}^\gb \,, P_{\gga\gd}] = -i \left (\gd^\gb_\gga P_{\ga\gd}
+\gd^\gb_\gd P_{\ga\gga}\right )\,,\qquad
[L_\ga{}^\gb \,, K^{\gga\gd}] =i \left (\gd^\gga_\ga K^{\gb\gd}+
\gd_\ga^\gd K^{\gb\gga}\right )\,,\qquad
\eee
\bee \nn
[P_{\ga\gb} \,, K^{\gga\gd}] = \f{i}{4}
\Big (\gd_\gb^\gga L_\ga{}^\gd + \gd_\ga^\gga L_\gb{}^\gd +
\gd_\ga^\gd L_\gb{}^\gga + \gd_\gb^\gd L_\ga{}^\gga \Big ) \,.
\eee
 $P_{\ga\gb}$ and $K^{\ga\gb}$ are generators of the
generalized translations and special conformal transformations.
The $gl_M$ algebra spanned by $L_\ga{}^\gb $
decomposes into the central subalgebra associated with the
generalized dilatation generator
$
D= L_\ga{}^\ga
$
 and the  $sl_M$ generalized  Lorentz generators
$
l_\ga{}^\gb = L_\ga{}^\gb -\f{1}{M}\delta^\gb_\ga  D\,.
$
Let us note that the coordinates $X^{\ga\gb}$ of $\M_M$ can be
interpreted as being dual to the generalized momenta $P_{\ga\gb}$
on the right hand side of the supersymmetry algebra
$\{Q_\ga , Q_\gb\}= P_{\ga\gb}$ which plays the key role in the
brane physics \cite{malg}.

The $Sp(8)$ invariant field equations considered in \cite{BHS,Mar} have the form
\be
\label{oscal} \Big (
\f{\p^2}{\p X^{\ga\gb} \p X^{\gga\gd}} - \f{\p^2}{\p X^{\ga\gga} \p
X^{\gb\gd}}\Big ) b(X) =0
\ee
for a scalar field $b(X)$ and
\be
\label{ofer} \f{\p}{\p X^{\ga\gb}} f_\gga(X)       -
\f{\p}{\p X^{\ga\gga}} f_\gb(X)      =0
\ee
for a svector field $f_\gb(X)$.
(We use the name ``svector" (symplectic vector)
to distinguish $f_\gb(X)$
from  vectors of the usual Lorentz algebra $o(d-1,1)$.
Note that svector fields obey the Fermi statistics \cite{Mar}).
For $M=2$, because antisymmetrization of any two-component indices
$\ga$ and $\gb$ is equivalent to their contraction with
the $2\times 2 $ symplectic form $\gvep^{\ga\gb}$, (\ref{oscal})
and (\ref{ofer}) coincide with the $3d$ massless Klein-Gordon and Dirac
equations, respectively. For $M=4$,  the equations
(\ref{oscal}) and (\ref{ofer}) in the generalized ten-dimensional
space-time $\M_4$ were shown in
\cite{Mar} to encode the infinite set of the usual
$4d$ equations of motion for massless  fields of all spins.

One interpretation of the extended matrix space-time
is that it is merely a  useful technical device like superspace
for supersymmetry. It is
interesting however to see what are specificities of the
dynamics in the generalized space-time $\M_M$  treated as a
physical space-time. In \cite{Mar} the dynamics in $\M_M$
described by the equations (\ref{oscal}) and (\ref{ofer})
was shown to be  consistent
with the principles of relativistic quantum field theory including
unitarity and microcausality. Most important difference compared to
the usual picture is that, because the system of equations
(\ref{oscal}) and (\ref{ofer}) is overdetermined, the true local phenomena
occur in a
submanifold $\gs$ called local Cauchy surface in \cite{Mar}
and identified with the usual space of Minkowski space-time.
The formulations of $Sp(2M)$ invariant systems
in terms of the generalized space-time $\M_M$
and usual space-time are equivalent and
complementary. The description in terms of $\M_M$ provides clear
geometric origin for the $Sp(2M)$ generalized conformal symmetry.
In particular it provides a geometric interpretation of the
electromagnetic duality transformations as particular generalized
Lorentz transformations. However, to define true local
fields, one has to resolve some constraints.
The description in terms of the Minkowski space-time,
that solves the latter problem, makes some of the symmetries
not manifest, namely, those that shift $\gs$.

It was argued in \cite{Mar} that it should be possible to
formulate different $Sp(2M)$ invariant equations in the same
space $\M_M$ associated with local Cauchy surfaces of different
dimensions. As a result, fields associated with different types
of $Sp(2M)$ invariant equations will visualize  $\M_M$ as Minkowski
space-times of different dimensions which may be interpreted as
different ``branes'' imbedded into the same generalized
space-time $\M_M$.  Our aim here is
to give  examples of such $Sp(2M)$ invariant equations
and to analyze their properties. We will see that elementary solutions
of the obtained equations will give a field-theoretical realization of
composites of  BPS ``preon'' states introduced in \cite{ban}.
 One of the observations of this paper is
that the dynamical equations for some of the rank 2  fields
are exactly of the form of
conservation conditions for ``conserved'' forms derived
in \cite{cur}, which give rise to the full set of
bilinear conserved charges in the rank 1
conformal higher spin theory. On the other hand, we will show that
the rank 2 system with respect to $sp(2M)$ is equivalent to the rank
1 system with respect to $sp(4M)$, which fact fits nicely the idea
of generalized AdS/CFT correspondence \cite{BHS}.
In particular, it implies that there are conserved currents
of degree $2^{p+1}$ built from the original rank 1 fields
in $\M_M$ to be integrated over branes of dimension
$2^p\, M$ to produce conserved charges independent of local
variations of branes.
(Recall that $\dim \M_M =\half M(M+1)$.)

We start in section 2 by summarizing some  general features
of the unfolded form of dynamical equations. An important
general fact emphasized here is that the space of solutions of
various partial differential equations invariant  under some symmetry
$g$ exhibits  a natural structure of associative algebra induced by the
tensor algebra of semiinfinite $g$-modules.
In section 3 dynamical equations associated with
 rank $r$ fields in $\M_M$ are considered.  The
full list of field equations in the rank 2 tensor product is given.
In particular it is shown that conserved currents,
constructed in \cite{cur}  from bilinears
of rank 1 fields in $\M_M$, satisfy some rank 2
field equations in $\M_M$. In section 4 it is shown that rank 2
fields in $\M_M$ are equivalent to the rank 1 fields in $\M_{2M}$.
In  section 5  generic solutions of the rank $2$
field equations are presented. It is argued that normalizable
solutions in $\M_M$ admit decomposition into positive and negative frequency
solutions and allow extension to the normalizable rank 1 solutions
in $\M_{2M}$. Section 6 contains
conclusions. Appendix contains some technicalities of the analysis
of the $\gs_-$ cohomology associated with dynamical
rank 1 and rank 2 fields and field equations in the framework of
the unfolded dynamics formalism.

\section{Unfolded Dynamics}
\label{unfol}

A natural approach to dynamical equations
of motion in the framework of the higher spin gauge theory, referred to
as ``unfolded formulation", consists of the reformulation of the dynamical
 equations in the form of some covariant constancy conditions \cite{Ann}.
Using this approach consistent gauge invariant
nonlinear higher spin equations of motion were found in
\cite{PV,4d} for the three dimensional and four dimensional theories.
It is particularly useful for revealing symmetries of dynamical equations,
as well as for the analysis of their dynamical content
(appropriate Cauchy data etc).
Since it  plays a key role in the analysis of this
paper let us summarize some of its  properties.

Let some dynamical system be reformulated in the form
\be \label{genequ}
(d +\go )C(X) = 0 \,,\qquad  d=d X^A\frac{\p}{\p X^A}\,,
\ee
where $C(X) $ denotes some (usually infinite) set of fields
taking values in a linear space $\V=V\times \Lambda$
where $V$  forms a module of some Lie algebra $g$ and
$\Lambda$ is the exterior (Grassmann) algebra on differentials $dX^A$.
In  other words, $C(X)$ is a section of the trivial vector
bundle $\B= R^d\times \V$   over the space-time base $R^d$
with the local coordinates $X^A$
$$
\begin{array}{rcl}
\V& \longrightarrow&\B\\
&&\downarrow\\
&& R^d \,.
\end{array}
$$
The 1-form $\go (X)=d X^A\go_A (X)$ is some fixed  connection of $g$
satisfying the flatness condition
\be
\label{R0}
d\go +\half [\go \,,\wedge  \go ]=0\,
\ee
(as usual, $[\,,]$ denotes the Lie product in $g$).

The equation (\ref{genequ}) is invariant under the global symmetry  $g$.
Actually, the system (\ref{genequ}) and (\ref{R0}) is invariant under
the infinitesimal gauge transformations
\bee \nn
\delta \go (X) = d \epsilon (X) + [\go (X) , \epsilon
(X)]\,,
\eee
\be
\label{glsym}
\delta C (X) = -\epsilon (X) C (X) \,,
\ee
where $\epsilon (X)$ is an arbitrary symmetry parameter taking
values in $g$. For a fixed $\go (X)$, there
is a leftover symmetry  with the parameter $\epsilon (X)$ satisfying
\be
\label{glpar}
\delta \go (X) \equiv d \epsilon (X) + [\go (X) , \epsilon (X)] =0\,.
\ee
This equation is consistent as a consequence of (\ref{R0}). Therefore, it
reconstructs the dependence of $\epsilon(X)$ on $X$ in terms of its values
$\epsilon(X_0)$ at any point of space-time $X_0$. The resulting
global symmetry algebra with the parameters $\epsilon(X_0)$ is $g$.
It is therefore enough to observe that some dynamical system
can be reformulated in the form (\ref{genequ}) with a flat connection $\go(X)$
taking values in some algebra $g$ that acts in $\V$
to reveal the global symmetry $g$ (\ref{glsym}) of the system (\ref{genequ}).
This approach is general since every free dynamical system can
be reformulated in the form (\ref{genequ}) by adding enough auxiliary
variables (nonlinear systems are described in
terms of an appropriate generalization associated with free differential
algebras; for more detail see e.g. \cite{Ann,Gol,Sah}).

In this paper we are interested in the particular
case of the equation (\ref{genequ}) of the form
\be \label{geneq}
(\D +\sigma_- )C (X)= 0 \,,
\ee
where the operators $\D$ and $\sigma_{-}$ have the properties
\be
\label{ds}
(\sigma_-)^2 =0\,,\qquad \D^2  =0\,,
\qquad \{ \D , \sigma_- \} =0\,,
\ee
which imply that the corresponding connection is flat.
Here $\D$ is a  covariant derivative for some $g^\prime  \subset g$  while
$\sigma_-$ is some vertical  operator  which acts
isomorphically on different fibers.
It is also assumed that there exists a grading operator $G$
diagonalizable in $V$, its spectrum  in $V$ is bounded from below
and \be \label{grad}
[G , \D ] = 0\,,\qquad [G,\sigma_- ] = - \sigma_-\,.  \ee
The exterior algebra $\E$ grading $F$, which counts a number of exterior
differentials $dX^A$, satisfies \be \label{gradex}
[F , \D ] = \D \,,\qquad [F ,\sigma_- ] =  \sigma_-\,.  \ee
The equations (\ref{geneq}) decompose into  independent subsystems for
$p-$forms $C(X)$  with different $p.$

Let us now recall some standard field theory
terminology. A field $A(x)$ is called auxiliary if it
is expressed by virtue of equations of motion as an
algebraic combination of a finite number of derivatives of
some other fields at the same point $x$. For example,
the equation
\be
\f{\p}{\p x} B(x) + A(x)=0
\ee
means that $A(x)$ can be chosen to be an auxiliary field.
All fields which are not auxiliary are called
dynamical fields. Those equations which express
auxiliary fields in terms of (derivatives of) some other
auxiliary fields and/or dynamical fields are called constraints.
By definition, upon elimination of all auxiliary fields by virtue of
constraints, dynamical fields satisfy some differential equations, called
dynamical equations. Dynamical equations can be trivial (no equations
at all) or nontrivial (some equations).

In other words, auxiliary fields are those which
can be eliminated from the
dynamical system in question without using nonlocal operators.
Note that the decomposition of a set of variables into auxiliary and
dynamical ones may be nonunique.  For example, in the
system of
equations $\f{\p}{\p x} B(x) + A(x)=0$ and $\f{\p}{\p x} A(x) + \alpha B(x)=0$,
where $\alpha$ is some dimensionful parameter, either
$A$ can be interpreted as an auxiliary field and $B$ can be interpreted as
the dynamical field or vice versa. The resulting two systems are equivalent
(dual) to each other. Some more ambiguity in
{ the definition of auxiliary fields is due to "triangular" field redefinitions}
\be \label{redef}
A\rightarrow A^\prime =A+ D(A,\phi)\,,\qquad \phi^\prime = \phi\,, \ee
where $A$ and $\phi$ denote auxiliary and dynamical fields, respectively, $D$ is
some differential operator, and  the inverse transform has analogous local form
$A^\prime \rightarrow A =A^\prime+ D^\prime(A^\prime,\phi)$ with some other
differential operator $D^\prime$. This ambiguity does not affect the dynamical
content of a system, namely the form of the dynamical field equations on the
dynamical fields. For the problems considered in this paper, where the
equations (\ref{geneq})  admit the grading $G$ (\ref{grad}),
the class of auxiliary field is defined uniquely modulo triangular
field redefinitions
(\ref{redef}) because auxiliary fields have higher $G-$grade compared
to the dynamical fields through derivatives of which they are expressed
by the constraints.

A useful observation is  \cite{SVsc,BHS} that
the  dynamical content of (\ref{geneq})
{ is encoded by the cohomology classes of the differential operator  $\sigma_-$           }
in $\V$, defined in the standard way $H^p(\sigma_- )= $
$Ker(\sigma_- )/Im (\sigma_- )\Big |_p$,
where $\Big |_p$ denotes the restriction to the subspace of $p$-forms
in $\V$. Note that $H^0 (\sigma_- ) =  Ker(\sigma_- ) \Big |_0$.

\newpage
{\bf Proposition 1}
\vspace{0.1cm}

Independent dynamical fields contained in the set of 0-forms $C$
take values  in $H^0 (\sigma_-)$.

\vspace{0.2cm}
This is because all fields taking values
in $V/H^0 (\sigma_-)$ (i.e., those with $\sigma_- C (X)\neq  0$)
are auxiliary being expressed via the
space-time derivatives of the dynamical fields by virtue
of (\ref{geneq}). Here we use the assumption that
the operator $\sigma_-$ is algebraic
in the space-time sense, i.e. it does not contain space-time derivatives.
Note that $H^0 (\sigma_-)$ is always non-zero because it
contains a nontrivial subspace of $V$ of minimal
grade.
 The ambiguity of the definition of the auxiliary
fields as elements of $V$ representing $V/H^0 (\sigma_-)$ is
irrelevant because it corresponds to the  field redefinitions
(\ref{redef}).

\vspace{0.2cm}
{\bf Proposition 2}
\vspace{0.1cm}

There are as many independent
differential equations on the dynamical 0-forms  contained in
(\ref{geneq}) as basis
elements of the cohomology group $H^1 (\sigma_- )$.
The differential equations  on grade $l$ dynamical fields,
associated with the grade $k$ elements of $H^1 (\sigma_- )$,
are of order $k+1-l$.

Indeed, consider the decomposition of the space of 0-forms $C$
into the direct sum of eigenspaces of $G$. Let a
field having definite eigenvalue $s_k$ of $G$ be
denoted $C_k$, $k= 0,1,2 \ldots$.
{ The equation (\ref{geneq})  decomposes into the infinite set of equations
\be
\label{geneqgr}
\sigma_- ( C_{k+1} )  + \D ( C_{k}) =0\,\qquad k= 0,1,2 \ldots.
\ee

Suppose that the dynamical
content of the equations (\ref{geneqgr}) with the eigenvalues}
$s_k$ with $k \leq k_q $ is found. Applying the operator $\D $
to the left hand side
{ of  the equations (\ref{geneqgr}) at  $k \leq k_q $ we obtain}
taking into account (\ref{ds})  that
\be
\label{sdc}
\sigma_- (\D ( C_{k_q+1} )) =0\,.
\ee
Therefore $\D ( C_{k_q +1} )$ is
$\sigma_-$ closed. If the group $H^1 (\sigma_- )$ is trivial
{ in the grade $k_q+1$ sector,
 any solution of (\ref{sdc}) can be written in}
the form $\D ( C_{k_q+1} ) = \sigma_- \tilde{C}_{k_q +2}$
for some  field $\tilde{C}_{k_q +2}$. This,
in turn, is equivalent to the statement that one can adjust
$C_{k_q +2}$ in such a way
that $\tilde{C}_{k_q +2} =0$ or, equivalently, that the part of the
{ equation (\ref{geneqgr}) of the grade $k_q+1$ is some constraint
that expresses $C_{k_q +2}$ in terms of the derivatives of $C_{k_q +1}$
(again, we  use the assumption that
the operator $\sigma_-$ is algebraic in the space-time sense). If
$H^1 (\sigma_-)$  is nontrivial  in the grade $k_q+1$ sector,
 a generic of (\ref{sdc}) can be written in the form}
\be
\label{d}
\D ( C_{k_q+1} )) = \sigma_- (\tilde{C}_{k_q+2}) + h_{k_q+1}\,,
\ee
where $h_{k_q+1}$ is some $\sigma_-$ closed element
which cannot be
absorbed into a redefinition of the auxiliary field $\tilde{C}_{k_q+2}$,
thus representing $H^1 (\sigma_- )$.
The equation (\ref{d}) expresses $\tilde{C}_{k_q+2}$ and  $ h_{k_q+1}$
in terms of derivatives of $C_{k_q+1}$ which is expressed in terms of
derivatives of the dynamical fields by the previously solved constraints.
The equation (\ref{d}) therefore is  the constraint for the auxiliary
fields $\tilde{C}_{k_q+2}$ and $h_{k_q+1}$.
The equation (\ref{geneq}) imposes the condition that $ h_{k_q+1} =0$.  It
is therefore equivalent to some differential equation
on the dynamical  fields.{  Thus, when $H^1 (\sigma_-)\neq 0$
in the grade $k_q+1$ sector,  (\ref{geneq}) not only
expresses the field $C_{k_q+2}$ in terms of  derivatives of
$C_{k_q+1}$ but also imposes some additional
 differential equations on
%$C_{k_q+1}$ and, so, on
the dynamical  fields}.

Note that if  $H^1 (\sigma_-)$ is zero, the equation
(\ref{geneq}) is equivalent to some  infinite set of
constraints which express all fields contained in
the decomposition of %here
$C(X)$ via derivatives of the dynamical fields.

The following comments are now in order.

{\bf Comment 1.}
\label{remark0}
When the fields $C(X)$ are $p$-forms, dynamical fields are associated
with $H^{p} (\gs_- )$ while nontrivial dynamical equations
are characterized by $H^{p+1} (\gs_- )$.

{ The conclusions also are true when
$\D$ ($(\D+\gs_-)^2=0$) also contains operators of negative grade,
since they do not affect the inductive analysis. }

{\bf Comment 2.}
\label{remark1}
Let us exchange the
roles of $\D$ and $\gs_-$ in the analysis
of the proposition 2. Suppose  that $\D$ has trivial
cohomology group $H^{1} (\D )$  and admits some grading operator $G^\prime$
which has spectrum bounded from below and satisfies $[G^\prime , \D] = -\D$.
The application to $\D$ the analysis analogous to that applied to $\gs_-$
shows that
there are no nontrivial consequences of the equation (\ref{geneq}) on the
 $0-$forms $C_0$.
  For example, this is true for the
de Rham differential $\D=d$ acting on the space $V^\prime$
of Taylor
expansions $C(X)$ in powers
of $X - X_0$ with $G^\prime$ counting a polynomial degree. In that case,
given  $C (X_0)=C_0$ for some $X_0$, the equation
(\ref{geneq})  reconstructs $C (X)$ uniquely  as
 the Taylor power series expansion at
$X=X_0$. This implies that the module $V^\prime$ can be identified with the
linear space of all space-time derivatives of $C(X)$ which are
allowed to be nonzero by the dynamical equations.
This interpretation is useful in several respects.
In particular it tells us that
if $V^\prime$ is spanned by unrestricted functions of some $p$ variables, the
Cauchy problem in the $X-$space should be formulated
in terms of $p-$dimensional Cauchy surfaces because the spaces
of functions parametrising the space of solutions have to be
of the same type. If, on the other hand, global consideration
changes $H^1 (\D )$ this
would imply that not every element of $V^\prime$ gives rise to some
global solution.

{\bf Comment 3.}
\label{remark2}
Suppose that $g$ admits a triple ${\bf Z}$ graded structure
$g = g_0 + g_- +g_+$ with Abelian subalgebras $g_\pm$. Let
$\gs_- = dX^A P_A$, where $P_A$ is some basis of $g_-$.
The dynamical fields $\gs_- (C)=0$ then identify with the
primary fields in $C(X)$ satisfying $P_A C(X)=0$\footnote{Note
that the role of ``translations" $P_A$
and ``special conformal" generators $K^A$ of $g_+$ acting in the
fiber is exchanged in the unfolded formulation
as compared to the standard induced representation
approach \cite{ind} in which primaries are defined directly
in the base manifold.}.
In other words, $H_0 (\gs_- )$
consists of singular vectors (i.e., vacua) of the irreducible $g-$submodules
      in $V$.  Obviously, any invariant submodule in $V$ gives rise
to a subsystem in (\ref{geneq}).
$H_0 (\gs_- )$
forms a $g_0-$module. In most interesting physical
applications $H_0 (\gs_- )$ decomposes into (may be infinite)
direct sum of finite-dimensional representations of $g_0$.

{\bf Comment 4.}
\label{remark3}
Suppose that $V_{1,2}=V_1\otimes V_2$. Let $C_1(X)\in V_1$
and $C_2(X)\in V_2$ solve (\ref{geneq}) with some operators
$\gs_{1-}$ and $\gs_{2-}$. Then
\be
\label{CCC}
C_{1,2} (X) = C_1 (X)\otimes C_2 (X)
\ee
solves (\ref{geneq}) with
\be
\label{gs12}
\gs_{1,2-} = \gs_{1-} \otimes Id +Id \otimes \gs_{2-}\,.
\ee
The independent  equations contained in
(\ref{geneq}) for $V_{1,2}$ with the $\gs_-$ operator (\ref{gs12})
are associated with
$H^1 (\gs_{1,2-})$ (which may be trivial, however).
As a result,
we see that the unfolded formulation of the dynamical equations
equips the variety of $g$-quasiinvariant partial differential
equations with the structure of associative algebra isomorphic
to the tensor algebra of semiinfinite $g$-modules.
{ Note that this} associative structure should not be confused with that of
the ring of  solutions of first order differential equations
because it maps solutions of
some set of (not necessarily first order) $g$-quasiinvariant
partial differential equations to solutions of some other
$g$-quasiinvariant equations.

One can consider  composite fields  of a particular symmetry type
associated with some irreducible representation of the
symmetric group $S_n$ acting on the $n$-fold tensor product of some
module $V$. In particular,
$C_{1,2} (X) = C (X)\otimes C(X)$ solves (\ref{geneq}) in the
symmetric tensor square of $V$.
For example, for the
model considered in this paper we will show  that
the formula (\ref{CCC}) produces conserved currents bilinear
in the original dynamical fields while
the new equations
are equivalent to the conservation conditions of
\cite{cur}.
Finding primaries in $C_{1,2} (X) $ (\ref{CCC}) turns
out to be equivalent to the construction of conserved currents
out of $C_1 (X)$ and $C_2 (X)$.
As  argued in section \ref{Solutions}, the formula (\ref{CCC}) provides
a basis for generalized $AdS/CFT$
correspondence with the composite dynamical fields
contained in $C_{1,2} (X) $
interpreted as elementary  fields in a larger ``bulk'' space.

In this paper we apply this general approach to the analysis
of the $Sp(2M)$ invariant dynamics.
The equations (\ref{oscal}) and (\ref{ofer}) were derived
in \cite{BHS} from the following unfolded system of equations
\be
\label{dydy}
\left(
dX^{\ga\gb} \f{\p}{\p X^{\ga\gb}} +
dX^{\ga\gb}\f{\p^2}{\p y^\ga \p y^\gb}\right ) C(y|X) =0\,,
\ee
where $y^\ga$ are some auxiliary commuting variables.
To make contact with (\ref{geneq})-(\ref{grad}) we set
\bee \nn
\D  =dX^{\ga\gb} \f{\p}{\p X^{\ga\gb}} ,\qquad
\sigma_- =dX^{\ga \gb}\frac{\ptl^2}{\ptl y^\ga\ptl {y}^\gb}\,,\qquad
G = \half {y}^\ga \f{\ptl}{\ptl {y}^\ga }.
\eee
The operators $\D$ and $\sigma_-$ act  on sections
of the trivial vector bundle $\B=\M_M\times \V$ over $\M_M$
with the local coordinates $X^{\ga \gb}$
$$
\begin{array}{rcl}
\V& \longrightarrow&\B\\
&&\downarrow\\
&& \M_M
\end{array}
$$
with the fiber $\V=V\times \Lambda$ , where
$V$ is the space of power series
$$
f(y) =
\sum_{n=0}^\infty
f_{\gga_1 \ldots \gga_n} (X)
y^{\gga_1}\ldots y^{\gga_n}
$$
 in $y^\ga$.
The $0-$form $C(y|X)$  in (\ref{dydy} ) is a section of $\B$.
The spectrum of the grading operator
$G$ is bounded from below because $G$
counts a half of the degree of a polynomial.
Let a field with definite eigenvalue $\half k$ of $G$ be
denoted $C_{\f{k}{2}} (x)$, $k= 0,1,2 \ldots$.
The equation (\ref{dydy}) decomposes into the infinite set of
equations\footnote{Let us note that
equations analogous to (\ref{dydy}) of the form
$
\left(
dX^{\ga\gb} \f{\p}{\p X^{\ga\gb}} +
dX^{\ga\gb}y_\ga y_\gb \right ) C(y|X) =0
$
were derived as
particular constraints in the world line model of \cite{BLS}.
These equations are equivalent to the equations (\ref{dydy})
for a class of problems which allow Fourier transform
in the twistor variables $y_\ga$ but are of little use for the
$\sigma_-$ cohomology
analysis of a form of different types of $sp(2M)$ invariant
partial differential equations we focus on
in this paper because the operator $dX^{\ga\gb}y_\ga y_\gb$
increases rather than decreases the grading $G$ in the space of
polynomials of $y_\ga$. As a result, the counterpart of the
infinite system of equations (\ref{set2}) resulting from the
expansion in powers of $y_\ga$ of the equations of \cite{BLS}
does not admit a meaningful interpretation in terms
of a nontrivial system of differential equations on a
finite set of dynamical fields supplemented with certain
constraints on an infinite set of auxiliary fields.}
\bee
\label{set2}
\frac{\ptl^2}{\ptl y^\ga\ptl {y}^\gb}
 C_{1} (y|X) =-
\f{\p}{\p X^{\ga\gb}} C_0(y|X)
                        ,\nn\\
\frac{\ptl^2}{\ptl y^\ga\ptl {y}^\gb}
 C_{3/2} (y|X) =-
\f{\p}{\p X^{\ga\gb}} C_{1/2} (y|X),\nn\\ \dots\quad\quad, \\
\frac{\ptl^2}{\ptl y^\ga\ptl {y}^\gb}
 C_{1+k/2}(y|X) =-
\f{\p}{\p X^{\ga\gb}} C_{k/2}(y|X) \nn \\ \nn \dots\quad\quad.
\eee
Obviously, for this case  $H^0 (\gs_- )$ is spanned by constants and
linear polynomials
\bee  \nn
C_0(y|X)  = b(X) , \quad
C_{1/2}(y|X)   =y^\ga f_\ga (X),
\eee
which give rise to the dynamical variables in the equations
(\ref{oscal}) and (\ref{ofer}) \cite{BHS}. In their turn, the equations
(\ref{oscal}) and (\ref{ofer}) are consequences of the
trivial identity
$
%\label{det1.5}
\frac{\ptl^2}{\ptl y^\ga\ptl {y}^\gb}
\frac{\ptl}{\ptl y^\gga} -
\frac{\ptl^2}{\ptl y^\gga\ptl {y}^\gb}
\frac{\ptl}{\ptl y^\ga}
\equiv 0,
$
from which  it follows by virtue of
(\ref{set2}) that
\bee \nn
%\label{dd11}
\f{\p}{\p X^{\ga\gb}}
\frac{\ptl}{\ptl y^{\gga}}
C_{k/2}(y|X)
-
\f{\p}{\p X^{\ga \gga}}
\frac{\ptl}{\ptl y^{\gb}}
C_{k/2}(y|X)=0
\eee
and
\bee \nn
%\label{dd2}
\f{\p}{\p X^{\ga\gb}}
\f{\p}{\p X^{\gga\gd}}
C_{k/2}(y|X)
-
\f{\p}{\p X^{\ga\gga}}
\f{\p}{\p X^{\gb\gd}}
C_{k/2}(y|X)=0.
\eee
Setting here $y=0$ one recovers (\ref{ofer}) and (\ref{oscal}),
respectively.
To make sure that the equations (\ref{oscal}) and (\ref{ofer}) encode
the whole dynamical content of the
equation (\ref{dydy}) one has to analyze
 $H^1 (\gs_-)$. In \cite{BHS} it was claimed that the
$\sigma_-$-closed 1-forms, which are not exact, are
\bee \label{H1}
\ e_{\gga \gd,\gb} y^\gb dX^{\gga \gd}\,, \qquad
e_{\gga \gd,\gb \ga} y^\gb y^\ga dX^{\gga \gd}\,,
\eee
where   $e_{\gga \gd,\gb}$ and $e_{\gga \gd,\gb \ga}$
are arbitrary tensors having symmetry properties of the
Young tableaux
\quad
\begin{picture}(13,12)(0,0)
%kruk 2*2  melkij
{\linethickness{.250mm}
\put(00,00){\line(1,0){05}}
\put(00,05){\line(1,0){10}}
\put(00,10){\line(1,0){10}}
\put(00,00){\line(0,1){10}}
\put(05,00){\line(0,1){10}}
\put(10,05){\line(0,1){05}}
}
\end{picture}
and
\quad
\begin{picture}(13,12)(0,0)
%okno 2*2  melkoe
{\linethickness{.250mm}
\put(00,10){\line(1,0){10}}
\put(00,05){\line(1,0){10}}
\put(00,00){\line(1,0){10}}
\put(00,00){\line(0,1){10}}
\put(05,00){\line(0,1){10}}
\put(10,00){\line(0,1){10}}
}
\end{picture}
, respectively\footnote{Let us note that tensors
(multisvectors) are classified here according to
 the representations the grade zero
group $GL_M$ which acts on homogeneous
polynomials in $y^\ga$. The grading operator $G$
is its central element. It is a generalization of the
dilatation in the usual conformal algebra.  In this paper
Young tableaux characterize representations of the
$GL_M$.} (i.e., $e_{\gga \gd,\gb} =e_{\gd \gga ,\gb} $,
$e_{\gga \gd,\gb \ga}=e_{\gd \gga,\gb \ga}=e_{\gga \gd,\ga \gb}$
and symmetrization over any three indices gives zero).
The explicit proof of this fact is given in Appendix.
The left hand sides of the equations
 (\ref{ofer}) and  (\ref{oscal}) form tensors of the
same symmetry types as the respective elements of $H^1 (\gs_-)$ in
(\ref{H1}). Thus, the equations (\ref{oscal}), (\ref{ofer})    form the
full system of dynamical equations of motion. All other
equations in (\ref{dydy})
either express auxiliary fields in terms of derivatives of
$b(X)$ and $f_\ga (X)$ or are consequences of these expressions.

The system (\ref{dydy}) provides an example of the
equation (\ref{genequ}) with the symmetry algebra
$g=sp(2M)$. The field $C(y|X)$ can be interpreted as
taking values in the Fock module ${F}$
generated by the creation operators
$y^\ga$, i.e.
\be
\label{Fock}
|C(y,X)\rangle = C(y,X)|0\rangle\,,\qquad \bar{y}_\ga   |0\rangle =0\,,
\ee
where
\bee \nn
[\bar{y}_\ga \,,y^\gb]=\delta_\ga{}^\gb \,,\qquad
[y^\ga \,,y^\gb]=0\,,\qquad
[\bar{y}_\ga \,,\bar{y}_\gb]=0\,.
\eee
The generators of $sp(2M)$ are realized as various
bilinears built from $y^\ga$ and $\bar{y}_\ga$,
\bee \nn
P_{\ga\gb} = \bar{y}_\ga \bar{y}_\gb\,,\qquad
K^{\ga\gb} = {y}^\ga {y}^\gb\,,\qquad
L_{\ga}{}^{\gb} = \half \{\bar{y}_\ga\,, {y}^\gb\}\,.
\eee
Also it is obvious that the infinite-dimensional algebra
of various polynomials in the oscillators $y^\ga$ and
$\bar{y}_\ga$ acts on the Fock module ${F}$. As a result,
the Weyl algebra generated by the
oscillators $y^\ga$ and $\bar{y}_\ga$ forms  global symmetry
of the system of equation (\ref{dydy}). This is the higher spin
conformal algebra discussed in \cite{FL,3d,BHS}.

The operator
$
\gs_- = dX^{\ga\gb} P_{\ga\gb}
$
provides a particular flat connection of $sp(2M)$
with the only nonzero part in the sector of generalized translations
in $sp(2M)$. Thus, according to Comment 3, dynamical fields are
identified with the primaries of $sp(2M)$ with respect to $P_{\ga\gb}$.

 As shown in \cite{BHS}, the $sp(2M)$ global symmetry transformations
(\ref{glsym}), (\ref{glpar}) of the dynamical fields are
\bee  \nn
%\label{db}
\delta b(X) &=&\Big ( \gvep^{\ga\gb} \f{\p}{\p X^{\ga\gb}}
+\half\gvep^{\ga}{}_{\ga}+2\gvep^{\ga}{}_{\gb} X^{\gb\gga}
\f{\p}{\p X^{\ga \gga}}\nn\\&-&
\gvep_{\ga\gb}\Big [\half X^{\ga\gb} + X^{\ga\gga}
X^{\gb{\eta}}\f{\p}{\p X^{\gga{\eta}}}\Big ] \Big )b(X)\,,
\eee
\bee   \nn
%\label{df}
\delta f_\gga(X)\ls\, &=&\ls\,\Big ( \gvep^{\ga\gb} \f{\p}{\p X^{\ga\gb}}
+\half\gvep^{\ga}{}_{\ga}+2\gvep^{\ga}{}_{\gb} X^{\gb\eta}
\f{\p}{\p X^{\ga
\eta}}\nn\\&-&
\gvep_{\ga\gb}\Big [ \half X^{\ga\gb} + X^{\ga\delta}
X^{\gb{\eta}}\f{\p}{\p X^{\delta{\eta}}}\Big ] \Big )f_\gga(X)
+\left (\gvep^{\gb}{}_\gga - \gvep_{\eta\gga}X^{\eta\gb} \right )f_\gb(X)  ,
\eee
where $\gvep^{\ga\gb}$, $\gvep^\ga{}_\gb$ and $\gvep_{\ga\gb}$ are
$X-$independent
parameters of generalized translations, Lorentz transformations along with
dilatations, and special conformal transformations, respectively.
These transformations can be extended to $OSp(1,2M)$ acting on the
supermultiplet formed by scalar $b(X)$ and svector $f_\ga (X)$ and to
extended conformal supersymmetry $OSp(L,2M)$ acting on the appropriate
sets of scalars and svectors \cite{BHS} (see  also \cite{Mar}).

Note that the Fock module (\ref{Fock}) used to describe classical
dynamics is not unitary. However, it is dual by virtue of
an appropriate Bogolyubov transform \cite{3d,BHS} to the
unitary Fock module  describing the space of quantum states
with the positive norm.

\section{Tensor product}

Analogous construction can be applied to the rank $r$ tensor product
of any number $r$ of the Fock representations of $sp(2M)$. The
generators of $sp(2M)$ then have a form of a sum of  bilinears
built from $r$ mutually commuting copies of  oscillators
\bee  \nn
[\bar{y}_{i\ga} \,,y_j^\gb]=\delta_{ij}\delta_\ga{}^\gb \,,\qquad
[y_i^\ga \,,y_j^\gb]=0\,,\qquad
[\bar{y}_{i\ga} \,,\bar{y}_{j\gb}]=0\,,\qquad i,j = 1\ldots r\,.
\eee
This construction is typical for the oscillator realization of
representations of symplectic algebras \cite{BG}.

Rank $r$ dynamics in ${\cal M}_M$ is described by a field $C(y|X)$
polynomial in the real variables $y_{i}^{\ga}$
\bee  \nn
C(y|X)=\sum_n
f_{\ga_1 \ldots \ga_{n}}^{i_1 \ldots i_{n}}(X)
y_{i_1}^{\ga_1}\cdots y_{i_n}^{\ga_{n}},
\eee
where
$\ga_j =1,\ldots,M$, $i_j=1,\ldots,r$ and
$X^{\gga \gb}=X^{\gb \gga}$ are real matrix coordinates of
${\cal M}_M$. The field $C(y|X)$ can be thought of as
taking values in the rank $r$ tensor product
$\underbrace{F\otimes\dots\otimes F}_{r}$ of the Fock
module ${F}$.
The  operator $\gs_-$ has the form
\be \label{sminusT}
\gs_-=\sum_{i=1}^r
dX^{\ga \gb}\frac{\ptl^2}{\ptl y^\ga_i\ptl {y}^\gb_i}\,,
\ee
so that the equation (\ref{geneq}) reads
\be
\label{dydyr}
\left(
dX^{\ga\gb} \f{\p}{\p X^{\ga\gb}} +
\sum_{i=1}^r
dX^{\ga \gb}\frac{\ptl^2}{\ptl y^\ga_i\ptl {y}^\gb_i}
\right)
C(y|X) =0\,.
\ee
According to the general argument of section \ref{unfol},  it
is invariant under the $sp(2M)$ global symmetry.

Let us note that the following equations are consequences
of (\ref{dydyr})
\be
\label{eqdetr}
\frac{\ptl}{\ptl X^{[\ga_1 \gb_1}}
\frac{\ptl}{\ptl X^{\ga_2 \gb_2}}
\cdots
\frac{\ptl}{\ptl X^{\ga_{r+1}] \gb_{r+1}}} C(y|X)=0\,
\ee
and
\be
\label{eqdetr1}
\frac{\ptl}{\ptl X^{[\ga_1 \gb_1}}
\frac{\ptl}{\ptl X^{\ga_2 \gb_2}}
\cdots
\frac{\ptl}{\ptl X^{\ga_{r} \gb_{r}}}
\frac{\ptl}{\ptl y^{\ga_{r+1}]}_i} C(y|X)=0\,,
\ee
where square brackets imply total antisymmetrization of the indices
$\ga_i$. The role of the equation (\ref{eqdetr})
is analogous  to that of the Klein-Gordon equation for the usual massless
fields. It is satisfied by every rank  $r$ field in $\M_M$.
The equation (\ref{eqdetr1}) is analogous to the Dirac equation.

According to the Comment 2 of Section 2, modules of solutions of
the rank $r$ equations in $\M_M$ are functions of $rM$ variables
$y_i^\ga$. As a result, the dimension of a ``local Cauchy bundle''
\cite{Mar} on which  initial data should be given to fix
a form of the solution everywhere in $\M_M$ is $rM$.

Let us now focus on the rank 2 case. Instead of
the real variables $y_1^\gga,y_2^\gd$ and fields $C(y_1 ,y_2 |X)$,
it is convenient to introduce complex variables
\be\label{compl}
\sqrt{2} z^\ga= y^\ga_1+i {y}_2^\ga\,,\qquad
\sqrt{2}{\bz}^\ga= y^\ga_1-i {y}_2^\ga
\ee
and field variables
\bee \nn
\C(z,{\bar z}|X)=
\sum
{c}_{\gga_1 \ldots \gga_k;\gd_1 \ldots \gd_n}(X)
z^{\gga_1} \cdots z^{\gga_k} {\bz}^{\gd_1} \cdots {\bz}^{\gd_n}.
%\equiv \sum c_{\gga(k);\gd(n)}z^{\gga(k)}\bz^{\gd(n)}.
\eee
In these terms
\bee \nn
\sigma_- =2 dX^{\ga \gb}
 \frac{\ptl^2}{\ptl z^\ga\ptl {\bz}^\gb}
\,
,\quad G = \half
\left ( z^\ga \f{\ptl}{\ptl z^\ga } +
\bar{z}^\ga \f{\ptl}{\ptl \bar{z}^\ga }\right )\,.
\eee
The equation  (\ref{dydyr}) now reads
\be
\label{dydy2z}
dX^{\ga\gb} \f{\p}{\p X^{\ga\gb}} +
2 dX^{\ga \gb}
\frac{\ptl^2}{\ptl z^\ga\ptl {\bz}^\gb}\,
C(z,\bar{z}|X) = 0\,.
\ee
The following simple consequences of the equation (\ref{dydy2z}) are
true
\be
\label{eqdet3}
\gvep^{\ga_1,\ga_2,\ga_3}
\frac{\ptl}{\ptl X^{\ga_1 \gb_1}}
\frac{\ptl}{\ptl X^{\ga_2 \gb_2}}
\frac{\ptl}{\ptl X^{\ga_3 \gb_3}}\C(z,\bar{z}|X)=0,
\ee
\be
\label{eqdet111}
\gvep^{\ga_1,\ga_2,\ga_3}
\frac{\ptl}{\ptl X^{\ga_1 \gb_1}}
\frac{\ptl}{\ptl z^{\ga_2}}
\frac{\ptl}{\ptl \bz^{\ga_3}}
\C(z,\bz|X)=0,
\ee
\be
\label{eqdet21}
\gvep^{\ga_1,\ga_2,\ga_3}
\frac{\ptl}{\ptl X^{\ga_1 \gb_1}}
\frac{\ptl}{\ptl X^{\ga_2 \gb_2}}
\frac{\ptl}{\ptl z^{\ga_3}}
\C(z,\bz|X)=0,
\ee
\be
\label{eqdet21b}
\gvep^{\ga_1,\ga_2,\ga_3}
\frac{\ptl}{\ptl X^{\ga_1 \gb_1}}
\frac{\ptl}{\ptl X^{\ga_2 \gb_2}}
\frac{\ptl}{\ptl {\bz}^{\ga_3}}
\C(z,\bz|X)=0,
\ee
\be
\label{eqdet211}
\gvep^{\ga_1,\ga_2}
\gvep^{\gb_1,\gb_2}
\frac{\ptl}{\ptl X^{\ga_1 \gb_1}}
\frac{\ptl}{\ptl z^{\ga_2}}
\frac{\ptl}{\ptl z^{\gb_2}}
\C(z,\bz|X)=0\,,
\ee
\be
\label{eqdet211b}
\gvep^{\ga_1,\ga_2}
\gvep^{\gb_1,\gb_2}
\frac{\ptl}{\ptl X^{\ga_1 \gb_1}}
\frac{\ptl}{\ptl {\bz}^{\ga_2}}
\frac{\ptl}{\ptl {\bz}^{\gb_2}}
\C(z,\bz|X)=0\,,
\ee
where $\gvep^{\ga_1,\ga_2,\ga_3}$ and $\gvep^{\ga_1,\ga_2}$ are
arbitrary totally antisymmetric tensors, introduced to impose
appropriate antisymmetrizations.

The equation (\ref{dydy2z}) decomposes into the infinite set of
subsystems associated with different integer
eigenvalues of the operator
${\cal H}$
\bee  \nn
{ {\cal H}} = h_z - h_{\bar{z}}\,,\qquad h_z =  z^\ga \f{\ptl}{\ptl
z^\ga}\,, \qquad h_{\bar{z}}=
\bar{z}^\ga \f{\ptl }{\ptl \bar{z}^\ga}\,,
\eee
which commutes with $\gs_-$.
Introducing notation $\C_{n,m}(z,\bz|X)$ for the
fields taking values in
the eigenspaces of the operators $h_z$ and $h_{\bz}$
\bee  \nn
h_z\C_{n,m}=n\,\C_{n,m} \,,\qquad  h_{\bz}\C_{n,m}=m\,\C_{n,m},\qquad
n,m\ge 0\,,
\eee
the  equation (\ref{dydy2z}) acquires the form
\bee   \nn
%\label{set3}
\left ( \frac{\ptl^2}{\ptl z^\ga\ptl {\bz}^\gb}
+\frac{\ptl^2}{\ptl \bz^\ga\ptl {z}^\gb}\right )
 \C_{n+1,m+1}(z,\bz|X) =-
\f{\p}{\p X^{\ga\gb}} \C_{n,m}(z,\bz|X)\,.
\eee

To find which fields among $\C(z,{\bar z}|X)$
are dynamical, one has to consider the cohomology group $H^0 (\gs_- ).$
A $\gs_-$ closed $0-$form $\C$  satisfies
\be
\label{kog0}
\left (\frac{\p^2}{\p
\bz^\ga \p z^\gb}+ \frac{\p^2}{\p z^\ga \p \bz^\gb}\right )\C=0 .
\ee

Being a $0-$form, $\C$ cannot be exact, i.e. $H^0(\gs_-) $
consists of all solutions of (\ref{kog0}).
As we show in Appendix,
$H^0(\gs_-) $ contains  the following  0-forms:
a constant $c(X)$, antisymmetric tensor
$c_{\ga, \gb}(X) z^{\ga}\bar{z}^{\gb} $,
 degree-$n>0$  analytic polynomials  $c_{\ga_1 \dots \ga_n }(X)
z^{\ga_1} \ldots z^{\ga_n }$ and
 degree-$n>0 $  anti-analytic polynomials
  ${\bar{ c}}_{\ga_1 \ldots \ga_n}(X)\bz^{\ga_1}\ldots \bz^{\ga_n}$.
\bee
\label{firsttab}
\begin{tabular}{|c| c| c|}% pervyj tabular
\hline
\rule[-3pt]{0pt}{23pt}
Dynamical 0-form & Young tableau & ${\cal H}$- eigenvalue \\
%\rule{0pt}{.1pt}&&\\
\hline
\rule[-3pt]{0pt}{23pt}
$c(X)$&$\, \bullet$&$0$\\
%\rule{0pt}{.1pt}&&\\
\hline
\rule[-3pt]{0pt}{23pt}
$c_{\ga, \gb}~(X) z^{\ga}\bar{z}^{\gb};\quad c_{\ga, \gb}=-c_{\gb, \ga} $&
\begin{picture}(09,13)
%stolbik 2
{\linethickness{.250mm}
\put(05,05){\line(1,0){05}}%
\put(05,10){\line(1,0){05}}%
\put(05,0){\line(1,0){05}}%
\put(05,0){\line(0,1){10}}%
\put(10,0.0){\line(0,1){10}}
}
\end{picture}
&$0$ \\
\hline
\rule[-3pt]{0pt}{23pt}
${c}_{\ga_1 \dots \ga_n}(X)z^{\ga_1} \cdots z^{\ga_n}$&
\begin{picture}(40,1)
%palka n
{\linethickness{.250mm}
\put(00,00){\line(1,0){40}}%
\put(00,05){\line(1,0){40}}%
\put(00,00){\line(0,1){05}}%
\put(05,00.0){\line(0,1){05}} \put(10,00.0){\line(0,1){05}}
\put(15,00.0){\line(0,1){05}} \put(20,00.0){\line(0,1){05}}
\put(25,00.0){\line(0,1){05}} \put(30,00.0){\line(0,1){05}}
\put(35,00.0){\line(0,1){05}} \put(40,00.0){\line(0,1){05}}
}
\put(21,6.2){\scriptsize  ${n}$}
\end{picture}
&$n>0$\\
%\rule{0pt}{.1pt}&&\\
\hline
\rule[-3pt]{0pt}{23pt}
${\bar c}_{\ga_1 \dots \ga_n}(X)\bz^{\ga_1} \cdots \bz^{\ga_n}$&
\begin{picture}(40,05)
%T_{\ga(n)}
%palka n   (8)
{\linethickness{.250mm}
\put(00,00){\line(1,0){40}}%
\put(00,05){\line(1,0){40}}%
\put(00,00){\line(0,1){05}}%
\put(05,00.0){\line(0,1){05}} \put(10,00.0){\line(0,1){05}}
\put(15,00.0){\line(0,1){05}} \put(20,00.0){\line(0,1){05}}
\put(25,00.0){\line(0,1){05}} \put(30,00.0){\line(0,1){05}}
\put(35,00.0){\line(0,1){05}} \put(40,00.0){\line(0,1){05}}
}
\put(20,6.2){{\scriptsize $n$}}
\end{picture}
&$-n<0$\\
\hline
\end{tabular}
\eee\\ \\
According to the general argument of section \ref{unfol},
all other components of $\C(z,{\bar z}| X)$ are
expressed by (\ref{dydy2z}) via  higher derivatives of
the dynamical fields.

Let us note that the obtained list is in the one-to-one correspondence
with the list of irreducible representations of $sp(2M)$ in the
tensor product of singletons \cite{G}.
This fact is not occasional, being a consequence of the
Comment 3  of section \ref{unfol}.
The precise matching between the decompositions of
the tensor products of the unitary singleton
modules and the non-unitary ones describing
classical fields in the unfolded formulation is due to
the  duality between quantum and classical descriptions
\cite{G1}, that was shown to have a form of a nonunitary
Bogolyubov transform  in \cite{SVsc} for  $M=2$  and then in \cite{BHS}
for $M=4$ and higher even $M$.

%here
Setting $z^\ga=\bar{z}^\ga =0 $ in (\ref{eqdet3})--(\ref{eqdet21b}),\,
 $\bar{z}^\ga =0 $ in (\ref{eqdet211}) and
 $z^\ga =0 $ in (\ref{eqdet211b})
 we obtain the following list of equations on the
dynamical fields (\ref{firsttab})
\bee
\label{qeqdet3}
\gvep^{\ga_1,\ga_2,\ga_3}
\frac{\ptl}{\ptl X^{\ga_1 \gb_1}}
\frac{\ptl}{\ptl X^{\ga_2 \gb_2}}
\frac{\ptl}{\ptl X^{\ga_3 \gb_3}}\, c(X)=&0& \qquad
\begin{picture}(25,20)(0,05)
%okno 2*3   melkij
{\linethickness{.250mm}
\put(00,05){\line(1,0){10}}%
\put(00,10){\line(1,0){10}}%
\put(00,15){\line(1,0){10}}%
\put(00,00){\line(1,0){10}}%
\put(00,00){\line(0,1){15}}%
\put(05,00.0){\line(0,1){15}}
\put(10,00.0){\line(0,1){15}}
}
\end{picture}, \qquad
\\
\nn \\
\label{qeq111}
\gvep^{\ga_1,\ga_2,\ga_3}
\frac{\ptl}{\ptl X^{\ga_1 \gb}} \, c_{\ga_2, \ga_3}~(X)=&0& \qquad
\begin{picture}(25,20)(0,05)
%kruk 2*3          \footnotesize
{\linethickness{.250mm}
\put(00,15){\line(1,0){10}}%
\put(00,10){\line(1,0){10}}%
\put(00,05){\line(1,0){05}}%
\put(00,00){\line(1,0){05}}%
\put(00,00){\line(0,1){15}}%
\put(05,00){\line(0,1){15}}
\put(10,10){\line(0,1){05}}
}
\end{picture},\qquad\qquad
\\
\nn \\
\label{qeq21}
\gvep^{\ga_1,\ga_2,\ga_3}
\frac{\ptl}{\ptl X^{\ga_1 \gb_1}}
\frac{\ptl}{\ptl X^{\ga_2 \gb_2}}\, c_{\ga_3}(X)=&0& \qquad
\begin{picture}(25,20)(0,05)
%tolstyi kruk 2*3  melkij
{\linethickness{.250mm}
\put(00,15){\line(1,0){10}}%
\put(00,10){\line(1,0){10}}%
\put(00,05){\line(1,0){10}}%
\put(00,00){\line(1,0){05}}%
\put(00,00){\line(0,1){15}}%
\put(05,00.0){\line(0,1){15}}
\put(10,05.0){\line(0,1){10}}
}
\end{picture},
\\
\nn \\
\label{qeq21b}
\gvep^{\ga_1,\ga_2,\ga_3}
\frac{\ptl}{\ptl X^{\ga_1 \gb_1}}
\frac{\ptl}{\ptl X^{\ga_2 \gb_2}}\, {\bar c}_{\ga_3}(X)=&0&  \qquad
\begin{picture}(25,20)(0,05)
%tolstyi kruk 2*3  melkij
{\linethickness{.250mm}
\put(00,15){\line(1,0){10}}%
\put(00,10){\line(1,0){10}}%
\put(00,05){\line(1,0){10}}%
\put(00,00){\line(1,0){05}}%
\put(00,00){\line(0,1){15}}%
\put(05,00.0){\line(0,1){15}}
\put(10,05.0){\line(0,1){10}}
}
\end{picture},
\\
\nn \\
\label{qeq211}
\frac{\ptl}{\ptl X^{\ga_1 \gb_1}}\, c_{\ga_2 \gb_2 \gga_1\dots\gga_{n-2}}(X)
-
\frac{\ptl}{\ptl X^{\ga_2 \gb_1}}\, c_{\ga_1 \gb_2 \gga_1\dots\gga_{n-2}}(X)& &
\nn\\
+
\frac{\ptl}{\ptl X^{\ga_2 \gb_2}}\, c_{\ga_1 \gb_1 \gga_1\dots\gga_{n-2}}(X)
-
\frac{\ptl}{\ptl X^{\ga_1 \gb_2}}\, c_{\ga_2 \gb_1 \gga_1\dots\gga_{n-2}}(X)
=&0&
\qquad
\begin{picture}(43,20)(0,05)
\put(22,11){\footnotesize n}    % melkij   dlinnyj dlinnyj kruk
{\linethickness{.250mm}
\put(00,05){\line(1,0){40}}%
\put(00,10){\line(1,0){40}}%
\put(00,00){\line(1,0){10}}%
\put(00,00){\line(0,1){10}}%
\put(05,00.0){\line(0,1){10}} \put(10,00.0){\line(0,1){10}}
\put(15,05.0){\line(0,1){05}} \put(20,05.0){\line(0,1){05}}
\put(25,05.0){\line(0,1){05}} \put(30,05.0){\line(0,1){05}}
\put(35,05.0){\line(0,1){05}} \put(40,05.0){\line(0,1){05}}
}
\end{picture} ,
\\
\nn \\
\label{qeq211b}
\frac{\ptl}{\ptl X^{\ga_1 \gb_1}}\, {\bar c}_{\ga_2 \gb_2
\gga_1\dots\gga_{n-2}}(X)
-
\frac{\ptl}{\ptl X^{\ga_2 \gb_1}}\, {\bar c}_{\ga_1 \gb_2
\gga_1\dots\gga_{n-2}}(X)& &\nn\\
+
\frac{\ptl}{\ptl X^{\ga_2 \gb_2}}\, {\bar c}_{\ga_1 \gb_1
\gga_1\dots\gga_{n-2}}(X)
-
\frac{\ptl}{\ptl X^{\ga_1 \gb_2}}\, {\bar c}_{\ga_2 \gb_1
\gga_1\dots\gga_{n-2}}(X)=&0&
\qquad
\begin{picture}(43,20)(0,05)
\put(22,11){\footnotesize n}    %
{\linethickness{.250mm}
\put(00,05){\line(1,0){40}}%
\put(00,10){\line(1,0){40}}%
\put(00,00){\line(1,0){10}}%
\put(00,00){\line(0,1){10}}%
\put(05,00.0){\line(0,1){10}} \put(10,00.0){\line(0,1){10}}
\put(15,05.0){\line(0,1){05}} \put(20,05.0){\line(0,1){05}}
\put(25,05.0){\line(0,1){05}} \put(30,05.0){\line(0,1){05}}
\put(35,05.0){\line(0,1){05}} \put(40,05.0){\line(0,1){05}}
}
\end{picture}  .
\eee  \\
Here  the Young tableaux describe the symmetry types of
the left hand sides of the field equations
and  $\gvep^{\ga_1,\ga_2,\ga_3}$ is an
arbitrary totally antisymmetric tensor.  Note that,
analogously to the fact that massless fields of any spin
in the Minkowski space-time satisfy the massless Klein-Gordon
equation, any solution  of the  equations
(\ref{qeq111})-(\ref{qeq211b}) satisfies (\ref{qeqdet3}).

To check whether this list of equations is complete,
one has to analyze $H^1 (\gs_-) $.
This analysis is analogous to that of $H^0 (\gs_- )$
but  somewhat more complicated. The details are given in Appendix.
The final result is that there are two $GL_M$ irreducible
representatives  of $H^1 (\gs_- )$
 with the ${\cal H}-$eigenvalue $h=0$ and just one
$GL_M$-irreducible class
for every $h\neq 0$:\\

\begin{tabular}{|c| c| c|}   %tretij tabular
\hline
\rule[-3pt]{0pt}{20pt}
 1-form & Young tableau & ${\cal H}$-eigenvalue\\
 \hline
$dX^{\gga_1 \gga_2}z^{\ga_1}z^{\ga_2}{\bar z}^{\gb_1}{\bar z}^{\gb_2}
\quad c_{\gga_1 \gga_2,\gb_1 \gb_2,\ga_1 \ga_2}(X)$&
\begin{picture}(25,20)
%okno 2*3   \footnotesize
{\linethickness{.250mm}
\put(00,05){\line(1,0){10}}%
\put(00,10){\line(1,0){10}}%
\put(00,15){\line(1,0){10}}%
\put(00,00){\line(1,0){10}}%
\put(00,00){\line(0,1){15}}%
\put(05,00.0){\line(0,1){15}}
\put(10,00.0){\line(0,1){15}}
}
\end{picture}&
$0$\\
 \hline
\rule[-10pt]{0pt}{25pt}
$dX^{\gga_1 \gga_2}
z^{\ga}\bar{z}^{\gb}
\quad c_{\gga_1 \gga_2,\ga,\gb}~(X)$&
\begin{picture}(25,20)
%kruk 2*3   \footnotesize
{\linethickness{.250mm}
\put(00,15){\line(1,0){10}}%
\put(00,10){\line(1,0){10}}%
\put(00,05){\line(1,0){05}}%
\put(00,00){\line(1,0){05}}%
\put(00,00){\line(0,1){15}}%
\put(05,00.0){\line(0,1){15}}
\put(10,10.0){\line(0,1){05}}
}
\end{picture}&
$0$\\
%\rule[-3pt]{0pt}{23pt}
\hline
\rule[-8pt]{0pt}{25pt}
$dX^{\gga_1 \gga_2}\bz^{\ga}{z}^{\gb_1}{z}^{\gb_2}
\quad c_{\gga_1 \gga_2,\gb_1 \gb_2,\ga }(X)$&
\begin{picture}(25,20)(0,05)
%tolstyi kruk 2*3                               \footnotesize
{\linethickness{.250mm}
\put(00,15){\line(1,0){10}}%
\put(00,10){\line(1,0){10}}%
\put(00,05){\line(1,0){10}}%
\put(00,00){\line(1,0){05}}%
\put(00,00){\line(0,1){15}}%
\put(05,00.0){\line(0,1){15}}
\put(10,05.0){\line(0,1){10}}
%\put(10,00.0){\line(0,1){10}}
}
\end{picture}&
$1$\\
\hline
\rule[-8pt]{0pt}{25pt}
$dX^{\gga_1 \gga_2}z^{\ga}{\bz}^{\gb_1}{\bz}^{\gb_2}
\quad {\bar c}_{\gga_1 \gga_2,\gb_1 \gb_2,\ga }(X)$&
\begin{picture}(25,20)(0,05)
%dlinnyi kruk n*2                                   \footnotesize
{\linethickness{.250mm}
\put(00,15){\line(1,0){10}}%
\put(00,10){\line(1,0){10}}%
\put(00,05){\line(1,0){10}}%
\put(00,00){\line(1,0){05}}%
\put(00,00){\line(0,1){15}}%
\put(05,00.0){\line(0,1){15}}
\put(10,05.0){\line(0,1){10}}
%\put(10,00.0){\line(0,1){10}}
}
\end{picture}&
$-1\quad$\\
\hline
\rule[-8pt]{0pt}{25pt}
$dX^{\gga_1 \gga_2}z^{\ga_1}\cdots z^{\ga_n}
\quad c_{\ga_1 \dots \ga_n, \gga_1 \gga_2}(X)$&
\begin{picture}(43,25)(0,05)
\put(22,11){{\footnotesize $n$}}
{\linethickness{.250mm}
\put(00,05){\line(1,0){40}}%
\put(00,10){\line(1,0){40}}%
\put(00,00){\line(1,0){10}}%
\put(00,00){\line(0,1){10}}%
\put(05,00.0){\line(0,1){10}} \put(10,00.0){\line(0,1){10}}
\put(15,05.0){\line(0,1){05}} \put(20,05.0){\line(0,1){05}}
\put(25,05.0){\line(0,1){05}} \put(30,05.0){\line(0,1){05}}
\put(35,05.0){\line(0,1){05}} \put(40,05.0){\line(0,1){05}}
}
\end{picture}&
$n\ge2$\\
%\rule[-3pt]{0pt}{23pt}
\hline
\rule[-8pt]{0pt}{25pt}
$dX^{\gga_1  \gga_2}\bz^{\ga_1}\cdots \bz^{\ga_n}
\quad {\bar c}_{\ga_1 \dots \ga_n, \gga_1  \gga_2}(X)$&
\begin{picture}(46,25)(0,05)
\put(22,11){\footnotesize  $n$}
{\linethickness{.250mm}
\put(00,05){\line(1,0){40}}%
\put(00,10){\line(1,0){40}}%
\put(00,00){\line(1,0){10}}%
\put(00,00){\line(0,1){10}}%
\put(05,00.0){\line(0,1){10}} \put(10,00.0){\line(0,1){10}}
\put(15,05.0){\line(0,1){05}} \put(20,05.0){\line(0,1){05}}
\put(25,05.0){\line(0,1){05}} \put(30,05.0){\line(0,1){05}}
\put(35,05.0){\line(0,1){05}} \put(40,05.0){\line(0,1){05}}
}
\end{picture}&
$-n\le-2$\\
\hline
\end{tabular}
\\[4pt]   \label{tabul3} \\ \\
{ An elementary analysis }
shows that the equations
(\ref{qeqdet3})-(\ref{qeq211b})
have correct symmetry properties and indeed result from
the general procedure described in the section \ref{unfol},
applied to the equation (\ref{dydy2z}).
Therefore  the list of the equations
(\ref{qeqdet3})-(\ref{qeq211b}) is complete. All other
equations in (\ref{dydy2z}) are either some constraints on the
auxiliary fields or their consequences.

Note, that the nontrivial dynamical equations associated with the
 elements of $H^1 (\gs_- )$
carrying some eigenvalue $h$ of $ {\cal H}$
are associated with  the dynamical fields carrying the
same $h$.

\section{Rank 2 -- rank 1 correspondence}

There are two basis relationships between systems of
ranks 1 and 2. The first one is that, as follows from
the comment 4 of section \ref{unfol},
bilinears of the rank 1 fields in $\M_M$
solve the rank 2 equations in $\M_M$. The second one is that the
rank 2 system in $\M_M$ turns out to be equivalent to
the rank 1 system in $\M_{2M}$.

The first property manifests itself in the fact that
the equations (\ref{qeq211}) and
(\ref{qeq211b}) have the  form of
the conservation condition for the generalized stress tensors
built in \cite{cur}, where it was derived from the requirement that
the generalized stress tensors should allow to build conserved
charges associated with the higher spin conformal symmetries
as integrals of some  closed  $M$-forms over $M$-dimensional
surfaces in $\M_M$. Namely, it was shown in \cite{cur} that
the  $M$-form
\bee   \nn
%\label{clo}
\Omega (\eta ) &=&
\epsilon_{\gamma_1 \ldots \gamma_M}
dX^{\gamma_1 \ga_1}\wedge \ldots \wedge dX^{\gamma_M \ga_M}\qquad\\ \nn
&{}& \eta_{\gb_1 \ldots
\gb_t}{}^{\ga_{M+1} \ldots \ga_{M+s}} X^{\ga_{M+s+1} \gb_1}\ldots
X^{\ga_{M+s+t}
\gb_t} T_{\ga_1 \ldots \ga_{M+s+t}}\,
\eee
is closed provided that the generalized stress tensor
$T_{\ga_1 \ldots \ga_{n}}$ satisfies the equation
(\ref{qeq211b}).
Here $\eta_{\gb_1 \ldots
\gb_t}{}^{\ga_{1} \ldots \ga_{s}}$ are some constants identified with
the parameters of the higher spin conformal
global symmetry generated by the conserved charges
\be
\label{charge}
Q(\eta) = \int_S \Omega (\eta)\,,
\ee
where $S$ is some $M$-dimensional surface in $\M_M$.
The explicit expression for
$
T(z) = \sum_{n=0}^\infty T_{\ga_1 \ldots \ga_{n}}z^{\ga_1}
\ldots z^{\ga_n}
$
in terms of bilinears of rank 1 fields, derived in \cite{cur}, is
\be
\label{TCC}
T^{k\,l}(z|X)= \C^k (z |X) \C^l (i z|X) \,.
\ee
(Here $k$ and $l$ are ``color'' indices taking an arbitrary number of
values).
It is now elementary to reveal its meaning in the context of the
comment 4 of section 2.
Indeed, once a  field $C^k(y|X)$
satisfies the equation (\ref{geneq}) with the rank 1
operator $\gs_-$, the rank 2 field of the form
\be
\label{CC}
C^{kl}(y_1 , y_2 |X)= C^k(y_1  |X) C^l(y_2 |X)
\ee
satisfies the equation (\ref{geneq}) with the rank 2
$\gs_-$ operator. The primaries in \\$C^k(y_1  |X) C^l(y_2 |X)$
that correspond to one-row Young tableaux are the generalized
stress tensors found in \cite{cur}.
Singling out the analytic and antianalytic parts of (\ref{CC})
in terms of the complex variables (\ref{compl}),
which correspond to the appropriate primaries, one
gets the formula (\ref{TCC}). (Let us note that the
usual conformal higher spin currents \cite{Ans,KVZ}
are also known \cite{Mikh} to be primaries
of the conformal group in the dual framework  of induced
representations.)

{}From the AdS/CFT correspondence perspective
the fact that the conservation conditions for currents built
from the rank 1 fields are particular equations of motion
in the rank  2 system
 suggests that, analogously to the case of usual conformal group
considered in \cite{KVZ}, there must be a correspondence
between bilinears built of rank 1 boundary fields and rank 2
bulk higher spin gauge fields in $\M_M$. Note that the
rank 2 fields $c$, $c_{\ga,\gb}$, and $c_{\ga (n)}$, $\bar{c}_{\ga (n)}$
(\ref{firsttab}) with $n <M$ are not associated with any
conserved currents. Presumably this implies that
they correspond to some non-gauge members of the
higher spin multiplets of a rank 2 theory, analogous to the
scalar and spinor fields in the $4d$ higher spin theory
\cite{Gol}.

 Let us now extend the rank 2 equations (\ref{dydyr}) to a larger system
\be
\label{1}
d X^{\ga \gb}\frac{\ptl}{\ptl X^{\ga \gb}}C\,+d X^{\ga \gb}\left(
\frac{\ptl^2}{\ptl y^\ga_1\ptl {y}_1^\gb}+ \frac{\ptl^2}{\ptl y^\ga_2\ptl
{y}_2^\gb}\right)C=0\,,
\ee
\be
\label{2}
d W^{\ga \gb}\frac{\ptl}{\ptl W^{\ga \gb}}\,C+d W^{\ga \gb} \left(
\frac{\ptl^2}{\ptl y^\ga_1\ptl {y}_1^\gb}- \frac{\ptl^2}{\ptl y^\ga_2\ptl
{y}_2^\gb}\right)C=0\,,
\ee
\be
\label{3}
 d Z^{\ga \gb}\frac{\ptl}{\ptl Z^{\ga \gb}}\,C + 2 d Z^{\ga \gb}
\frac{ \ptl^2}{\ptl y^\ga_1\ptl {y}_2^\gb}
C\,=0
\ee
with symmetric $W^{\ga \gb}$ and arbitrary  $Z^{\ga \gb}$.
This extension is consistent
(i.e., the corresponding connection is flat) and contains the original
rank 2 system (\ref{dydyr}) as a part
(\ref{1}). On the other hand, it is
nothing but the rank 1 system (\ref{dydy})
with the doubled number of auxiliary variables
$y^\Omega = (y_{1}^\ga , y_2^\gb)$. The consistency of
the system (\ref{1})-(\ref{3}) implies
that every solution  of (\ref{1}) is extended by virtue of
(\ref{2}) and (\ref{3})
to some solution of the whole system  because the equations
(\ref{2}), (\ref{3}) just reconstruct the dependence on the coordinates
$W$ and $Z$ for a given value of $C(y_{1,2}|X,W,Z)|_{W=Z=0}$.
Therefore, the rank 2 system in $\M_M$ is promoted to the rank 1 system in
$\M_{2M}$ in such a way, that every solution in the original rank 2
system
is promoted to some solution of the rank 1 system in $\M_{2M}$
and vise versa.

Among other things, this implies
that there exist conserved currents, built from the rank 2 fields
in $\M_M$, to be integrated over $2M$-dimensional surfaces in $\M_{2M}$
to produce conserved charges.
Choosing a $2M$ integration surface to belong to $\M_M$
(recall that $dim \M_M = \half M(M+1)$) one
gets generalized stress tensors constructed from rank 2 fields.
Substituting the conserved currents built from
bilinears of the rank 1 fields in $\M_M$ in place of the rank 2 fields,
one finds conserved charges being of fourth order in the original rank 1 fields
in $\M_M$, which are associated with the appropriate on-mass-shell closed $2M$
forms. That this process can be continued suggests that
there is a chain of dualities among various rank $2^p$ theories
by means of the rank doubling via further products analogous to
(\ref{CC}). This conclusion is consistent with the conjecture on
the existence of
an infinite chain of $AdS/CFT$ type dualities in higher spin
theories suggested in
\cite{BHS}\footnote{To avoid misunderstandings, let us note that this
kind of dualities was conjectured in \cite{BHS} to be true
for the chain of theories containing all higher spin massless
fields, which are different from the particular
reduced model associated with N=4 SYM theory for which no
infinite chain of dualities is expected.}.
Higher nonlinear combinations of the rank 1 fields should be
associated with higher rank gauge fields in $\M_M$.

\section{Solutions}\label{Solutions}

Let us analyze  solutions of the equations
(\ref{qeqdet3})-(\ref{qeq211b}). For
$ c(X)={c^{\prime}}\exp i k_{\ga \gb} X^{\ga \gb}$,
 the equation (\ref{qeqdet3}) requires
\be
\label{det}
\gvep^{\ga, \gb, \gd}
k_{\gga \ga} k_{\gn \gb} k_{\gm \gd}=0
\ee
for any totally antisymmetric $\gvep^{\ga, \gb, \gd}$.
This is solved by any  rank $r\le2$  matrix $k_{\ga \gb}$.
Indeed, any symmetric matrix
 $k_{\ga \gb}$ can be diagonalized by a $GL_M$ transformation.
{}From (\ref{det}) it follows that
the product of any three eigenvalues  of $k_{\ga \gb}$ is equal to zero.
So at most two of them can be nonzero. For this case one can write
\be
\label{spek}
k_{\ga \gb}=a_{i\,j}\xi_\ga^i\xi_\gb^j
\ee
with  two arbitrary vectors $\gx_{\ga}^i$ and
some symmetric form  $a_{i\,j}$. Eq. (\ref{spek}) provides
a general solution of (\ref{det}).

Because any solution of  the equations of
motion (\ref{qeqdet3})-(\ref{qeq211b})
satisfies (\ref{qeqdet3}), the respective harmonic solutions have a
 form
\bee
\label{genr}
{c_{...}}(X) = {c_{...}^{\prime}}(\xi)
 \exp i a_{i\,j}\xi_\ga^i\xi_\gb^j  X^{\ga \gb}\,. \eee
Specifically, the equations  (\ref{qeq111})-(\ref{qeq211b}) require
\bee \nn
{ c}_{\gga,\gd}(X)= &
 A^\prime(\xi )\, (\gx_{\gd}^1\gx^2_{\gga}-
\gx_{\gga}^1\gx^2_{\gd})
\exp (i a_{i\,j}\xi_\ga^i\xi_\gb^j  X^{\ga \gb})
 \qquad& \forall  A^\prime,\\  \nn
{c}_{\gga_1\dots\gga_n}(X)= &
A_{i_1\dots i_n}(\xi )\,\gx_{\gga_1}^{i_1}\dots\gx_{\gga_n}^{i_n}
(\exp i a_{i\,j}\xi_\ga^i\xi_\gb^j  X^{\ga \gb}) &
\forall n\ge0 ,\\ \nn
{\bar c}_{\gga_1\dots\gga_n}(X)=&
A_{i_1\dots i_n} (\xi )\,\gx_{\gga_1}^{i_1}\dots\gx_{\gga_n}^{i_n}
(\exp i a_{i\,j}\xi_\ga^i\xi_\gb^j  X^{\ga \gb}) &
\forall n\ge0,
\eee
where $A_{i_1\dots i_n}$  is an arbitrary tensor
traceless with respect to
the form $a^{ij}=\gvep^{i m}\gvep^{j k}a_{mk}$
$(\gvep^{ij}=-\gvep^{j i}, \gvep^{1 2}=1) $
\be     \label{trassless}
a^{i\,k} A_{i k j_3\dots j_{n}} =0\,.
\ee
Indeed, taking into account that the indices $i,j,k,\ldots$ take two values,
that allows one to replace any
antisymmetrization by the epsilon symbol
$\gvep_{i j}$,
it is  easy to see that the condition
$
a_{i\,k} A_{j_1 j_2\dots j_{n}} - a_{j_1\,k} A_{i j_2\dots j_{n}} -
a_{i\,j_2} A_{j_1 k j_3 \dots j_{n}} + a_{j_1\,j_2} A_{i k j_3\dots j_{n}}=0\,,
$
which follows from (\ref{qeq211b}), is equivalent to
(\ref{trassless}).

Let us now consider the general case of an arbitrary rank.
As mentioned in Section 3,
%Although we still do not have a full list of all higher rank
%$Sp(2M)$ invariant equation we know that
the scalar equation (\ref{eqdetr}) is satisfied by any solution.
This fixes the rank $r$ spectral condition for a particular harmonic
in the form
\bee \nn
%\label{detttt}
\gvep^{\ga_1, \dots, \ga_{r+1}}
k_{\gga_1 \ga_1}\cdots k_{\gga_{r+1} \ga_{r+1}} =0\,
\eee
for any totally antisymmetric parameter
$\gvep^{\ga_1, \dots, \ga_{r+1}}$.
This condition implies that the rank of the wave ``vector''
$k_{\ga\gb}$ cannot exceed $r$,  i.e. $k_{\ga\gb}$ is
described by the same formula (\ref{spek}),
in which the indices $i,j,\ldots $ take $r$ values.
Let us note that this formula has the same form as the
 $r$ ``preon'' representation \cite{ban} for momenta
in the brane models with partially broken
supersymmetries in the generalized space-time with
``central charge'' coordinates. The harmonic solutions in the
higher rank models in  $\M_M$, considered in this paper,
 therefore provide the field-theoretical realization of the
 BPS states with fractional supersymmetry in the
higher spin theories.

Generic solution of the scalar field equation (\ref{eqdetr})
can be represented in the form (\ref{genr}) with $i,j = 1\ldots r$.
By a $GL_r$ linear transformation
$
\xi^i_\ga \to A^i{}_j \xi^j_\ga
$
one can transform  $a_{i\,j}$ to one of the
canonical forms $a^{(p\,q)}_{i\,j}$,
\bee \nn      %matrica blochno diagonalnja, p 1 ,q -1
\beee{c r c}
&&
\beee{r r}
%\mbox{{ p}}
\mbox{{\scriptsize p}}
&\mbox{{\scriptsize q}}
\eeee
\\
a^{(p\,q)}  =&
\begin{tabular}{r} \mbox{{\scriptsize p\,}\,}\\\mbox{{\scriptsize q}\,}
\end{tabular}&
\left(
\beee{r r}
I&0\\0&-I
\eeee
\right)
\eeee
\eee
with $p+q \leq r$ (assuming that the rest elements of
$a^{p\,q}_{i\,j}$ vanish if $p+q <r$).
As a result, we have
\be
\label{Cdec}
C(X) =\sum_{p+q=r} C^{(p\,q) }(X)
\ee
where
\be
\label{p,q}
C^{(p\,q)}(X)=
\int  d^{M} \xi {f}^{(p\,q)}(\xi )
\exp(ia^{p\,q}_{i\,j}\xi_\ga^i\xi_\gb^j X^{\ga\gb} )\,.
\ee
Note
that the degenerate cases with $p+q < r$ are
described by  the formula (\ref{p,q}) with $p+q =r$
in which an appropriate $\delta$ - functional measure is
included into
$f^{(p\,q)}(\xi )$,
which
requires those components of $\xi^i_\ga$, which are null vectors
of  $a^{p\,q}_{i\,j}$, to vanish.

The  expression in the exponential (\ref{p,q})
is invariant under
$O(p,q)$ rotations acting on the indices $i,j$, which
leave the metric $a^{p\,q}_{i\,j}$ invariant. This means
that the ``Fourier components"
${f}^{(p\,q)}(\xi^i_\ga )$ are defined on $R^{rM}/O(p,q)$.
Note that analogous phenomenon takes place in the rank 1 system
of \cite{Mar} with the group $O(1)$ = $Z_2$ in place of $O(p,q)$.
A new feature of the rank $r>1 $ case is that it allows
noncompact situations with $pq \neq 0$. In this case a volume
of a typical orbit of $O(p,q)$ is  infinite and the expression
(\ref{p,q}) requires insertion of an appropriate delta-function
\be
\label{FP}
{f}^{(p\,q)}(\xi )=\delta (\chi (\xi ))\tilde{f}^{(p\,q)}(\xi)\,,
\ee
where $\chi(\xi )$ is some  gauge function
which fixes a representative of an orbit of $O(p,q)$.

Analogously to what was shown in \cite{Mar} for the rank 1 system,
the solutions $C^{(r\,0)}$ and $C^{(0\,r)}$ describe positive
and negative frequency modes to be associated with the creation and
annihilation operators upon quantization.  The solutions  $C^{(p\,q)}$
with $pq\neq 0$ do not allow
a decomposition into  positive and negative frequency parts and
thus
do not allow for a consistent quantization. The group-theoretical
meaning of the decomposition (\ref{Cdec}) is clear.
Let us for simplicity consider the rank 2 case. The solutions
of the equations (\ref{qeqdet3})-(\ref{qeq211b})  correspond to the
tensor product
of the solutions of the equations considered in \cite{Mar}.
Let the latter
be formally denoted  $b(X) = b^+ (X) +b^- (X)$,
where $b^+$ and $b^-$
are mutually conjugated positive- and negative-frequency parts.
Quantization identifies $b^+(X)$ with the unitary left module $U$ of
single-particle quantum states. $b^-(X)$ is associated with the conjugated
 right module
$\bar{U}$.  Then, the spaces of solutions $C^{(2\,0)},C^{(0\,2)}$
and $C^{(1\,1)}$ are
associated, respectively, with $ U \otimes U$, $ \bar{U}
\otimes \bar{U}$ and ${U} \otimes \bar{U}$.
If we would consider the quantum picture, by tensoring the
single-particle states we would only have
$U\otimes U$ and its conjugate. This suggests that the modes with
$pq \neq 0$ must be irrelevant.
Let us argue that, indeed, these modes are unstable and can
be ruled out
by requiring a solution to be normalizable with respect
to a natural norm.

 The equations (\ref{1})-(\ref{3})  show that every
solution of the rank  2 system can be
promoted to a solution of some rank  1 system in the larger space $\M_{2M}$.
It is not guaranteed however that
an oscillating solution in $\M_M$ remains oscillating in $\M_{2M}$.
For the rank 2 system, the extension  of a harmonic
solution to $\M_{2M}$ is
\be
\label{++}
C^{(20)}=
\int  d^{2M} \xi f^{(20)}(\xi )
\exp i((\xi_\ga^1\xi_\gb^1 +\xi_\ga^2\xi_\gb^2) X^{\ga\gb} +
(\xi_\ga^1\xi_\gb^1 - \xi_\ga^2\xi_\gb^2) W^{\ga\gb}
+2 \xi_\ga^1\xi_\gb^2 Z^{\ga\gb})
\,,
\ee
\be
C^{(11)}=
\int  d^{2M} \xi f^{(11)}(\xi )
\exp i((\xi_\ga^1\xi_\gb^1 -\xi_\ga^2\xi_\gb^2) X^{\ga\gb} +
(\xi_\ga^1\xi_\gb^1 + \xi_\ga^2\xi_\gb^2) W^{\ga\gb}
+2i\xi_\ga^1\xi_\gb^2 Z^{\ga\gb})
\,,
\ee
\be
\label{--}
C^{(02)}=
\int  d^{2M} \xi f^{(02)}(\xi )
\exp i((-\xi_\ga^1\xi_\gb^1 -\xi_\ga^2\xi_\gb^2) X^{\ga\gb} -
(\xi_\ga^1\xi_\gb^1 - \xi_\ga^2\xi_\gb^2) W^{\ga\gb}
-2 \xi_\ga^1\xi_\gb^2 Z^{\ga\gb})
\,.
\ee
We see that the extensions associated with
$C^{(2\,0)}$ and $C^{(0\,2)}$ are still oscillating in the extended
space $\M_{2M}$
while that associated with $C^{(1\,1)}$
exponentially grows along some of the
directions in $\M_{2M}$, thus exhibiting instability.
As observed in \cite{cur}, the space of solutions of the
rank 1
equation (\ref{dydy}) admits an invariant form $B(C,C)$
defined in terms of the conserved charge (\ref{charge}) as
$B(C,C) = Q(\eta)\Big |_{\eta =1}$. It has a form of an
integral over an  arbitrary
space-like $M$-dimensional surface $E$ in $\M_M$ and
is independent of  local variations
of $E $  provided that $C$ satisfies the field equations
 \footnote{Recall that the initial data problem
in $\M_M$ for the equations (\ref{oscal}) and (\ref{ofer}) is given on
a $M$-dimensional ``local Cauchy bundle" \cite{Mar}
because solutions are parametrized
 by functions of $M$ variables $\xi_\ga $.}.
Using this construction for the rank 1 solution
in $\M_{2M}$ generated from the rank 2 solution in $\M_M$
according to  (\ref{++})-(\ref{--}), one obtains the
invariant form on  rank 2 solutions.
Choosing an integration
surface in $\M_{2M}$ in such a way that some of the directions are
included along which a solution blows up, one can see that for such
solutions the norm gets  infinite. To make this argument
complete it remains to prove that such a rotation of
the integration surface can be obtained by a continuous deformation.
It is simpler, however, to use the Fourier transformed
representation for the norm in $\M_{2M}$ that was shown in \cite{cur}
to have a form
$
B(C,C)=\int d^{2M} \xi f(\xi) f(\xi)\,.
$
Inserting here (\ref{FP}) one finds that the delta-function gets
squared producing an infinite factor $\delta(0)$ originating from the
infinite volume of the orbit of $O(p,q)$. This factor
is finite only for the compact cases of $O(r,0)$ and $O(0,r)$
associated with the true positive- and negative-frequency solutions
that allow consistent quantization.
Thus, the modes $C^{(p\,q)}$ with $pq\neq 0$ are ruled out as
non-normalizable very much as the exponentially growing modes with
imaginary $k_{\ga\gb}$ in the rank 1 case, which are formally
allowed by the field equations.

The  normalizable solutions correspond
only to purely positive- or negative- frequency modes which,
by analogy with the analysis of  \cite{Mar}, are to be
associated with the single-particle spaces of quantum states
and their conjugates by virtue of quantization. In fact, the
norm $B(C^{(0\,r)} ,C^{(r\,0)})$ is the usual Fock space norm for these
quantum states \cite{Mar}. Thus, the normalizable
sector of the rank 2 system in $\M_M$ is equivalent to the
 normalizable rank 1 system in $\M_{2M}$.

\section{Conclusions}

In this paper the new $sp(2M)$ invariant equations of
motion, which describe propagation in $\half M(M+1)$-dimensional
space-time $\M_M$ with matrix coordinates
$X^{\ga\gb}$ ($\ga,\gb = 1\ldots M$), are derived.
The idea of the derivation  is based on the
unfolded formulation of the dynamical equations in the form of
covariant constancy equations imposed on the fields taking
values in some
module $V$ of the chosen symmetry algebra $g$. In \cite{BHS,Mar}
this scheme was applied to the Fock  module $V={F}$
for the oscillator representation of $sp(2M)$.
The resulting equations (\ref{oscal}) and (\ref{ofer}) were shown
to describe matter localized on a $M$-dimensional local Cauchy
bundle being some limiting $M$-dimensional surface in $\M_M$.
The equations formulated in this paper result from $V={F}\otimes {F}$
and, more generally,
 $\underbrace{{F}\otimes\dots\otimes {F}}_{r}$.

The list of various types of independent fields in the
proposed equations is shown to be in the one-to-one correspondence with
the content of the decomposition of the tensor product of a number of unitary
Fock  $sp(2M)-$modules $U$            into irreducible submodules. This fact
is in accordance with the
Bogolyubov transform type duality \cite{3d,BHS}
between the unitary modules of
single-particle quantum states and non-unitary  modules which appear in
the unfolded formulation of the classical field equations.

The space of solutions is parametrized by functions
of $rM$ variables. As a result, the local Cauchy bundles $E$
associated with the proposed field equations have dimension $rM$.
Note that as explained in
\cite{Mar}, the dimension of space $\gs$ where true localization of
events is possible, which is identified with the base manifold
of the local Cauchy bundle $E$, is generically lower
than the dimension of $E$.
 For example, for the lowest values of $M=2,4,8$, the
rank 1 local Cauchy bundles $E$
 have the structure $E=\gs\times S$ with the base manifolds
(local Cauchy surfaces)
$\gs=R^2, R^3, R^5$ as the physical spaces and
fiber compact manifolds $S= Z_2, S^1, SU(2) $,
respectively. This corresponds to { three-, four-
  and six-dimensional } Minkowski space-times.
The fibers $Z_2$, $S^1$ and $SU(2)$ give rise to some spin degrees
of freedom.  In particular, for the case of $M=2$, the modes of $Z_2$
are  $3d$ massless scalar and spinor. For the case of $M=4$, modes of
 $S^1$ give rise to the infinite tower of spins
in the $4-$dimensional Minkowski space-time \cite{Mar}.
One conclusion of this paper is that different local equations formulated
in the same space $\M_M$ may visualize it differently via its
subspaces of different dimensions. Moreover, it is shown that
the space-time $\M_M$ visualized through the rank two equations
is physically undistinguishable from the space-time $\M_{2M}$
visualized through the rank 1 system. This suggests that
it may be  enough to study the rank 1 systems in $\M_M$ with
various $M$.

Most of the rank 2 equations of motions obtained in this paper
have the form of the conservation conditions that were shown
in \cite{cur}  to give rise to the conserved charges, {defined as integrals
over $M-$dimensional surfaces in $\M_M$.}
% of therank 1 system.

This result  suggests the generalized $AdS/CFT$
correspondence  between
fields in rank 2 models and currents in rank 1 models. Remarkably,
the chain of correspondences can be continued to all rank $2^p$
models so that fields in the $2^p$ model correspond to bilinear
currents in the $2^{p-1}$ model,
 quartic currents in the
$2^{p-2}$ model etc. This result supports the
conjecture of \cite{BHS} that the full higher spin models may
exhibit infinite chains of $AdS/CFT$ dualities.
It is tempting to speculate
that there are two most
natural options. First is that the fundamental theory is the
one with $M=2$ which is the usual $3d$ conformal theory
or even with $M=1$\footnote{This case is degenerate.
One can speculate that it corresponds to left or right movers
of a  $2d$ conformal theory. Although there are no
nontrivial equations imposed on the dynamical fields in this case
one can formally apply the general arguments of this paper
to this system as well.}. An opposite
extreme option is that of $M= \infty$. The usual models,
like for example $4d$ higher spin theories,
may result from an appropriate
breakdown of $sp(\infty)$ down to, say, $sp(8)$.
The $sp(\infty)$ case has a good chance to be related to a higher spin
formulation of the superstring theory.
It may happen,
however, that the two seemingly opposite
options are not that different because
of the conjectured infinite chain of dualities.

The results of this paper suggest a new
dynamical mechanism for realization of branes of different dimensions
that live in the same generalized space-time $\M_M$.
The equations suggested in this paper provide a field-theoretical
realization of the ``preon'' construction
for BPS states suggested in \cite{ban}.
In a nonlinear version of the higher spin theory in the
generalized space-time
to be developed, fields associated with  branes of different dimensions
are expected to interact to each other.
Hopefully this will eventually lead to
the microscopic theory of branes.

\section*{Acknowledgments}
This research was supported in part by MIUR-COFIN contract 2001-025492,
INTAS, Grant No.99-01-590,
INTAS, Grant No.00-259 and the RFBR Grant No.02-02-17067.

\newcounter{appendix}
\setcounter{appendix}{1}
\renewcommand{\theequation}{\Alph{appendix}.\arabic{equation}}

%\newpage
\section*{Appendix. $\gs_-$
cohomology}

Let us first recall some elementary facts relevant to the
cohomology analysis. Let a linear operator $\Omega$
act in some linear space $\V$ and satisfy
$\Omega^2=0$. By definition, $H (\Omega )= \ker \,\Omega  /Im\, \Omega $
is the cohomology space. Let $\Omega^*$ be some other nilpotent
operator, $(\Omega^*)^2=0$. Then the operator
\be
\label{l1}
\Delta=\{ \Omega, \Omega^*\, \}
\ee
satisfies
\be
[\Omega , \Delta ] = [\Omega^*\,, \Delta ] =0\,.
\ee
{}From (\ref{l1}) it follows that
 $\Delta \ker \,\Omega \subset {\mbox Im} \,\Omega$. Therefore,
$H(\Omega )\subset \ker\, \Omega / \Delta ({\ker \,\Omega})$. Suppose
now that $\V$ is a Hilbert space in which
$\Omega^*\,$ and $\Omega$ are conjugated and that the operator $\Delta$
is quasifinite-dimensional, i.e. $\V=\sum \oplus \V_\ga$
with finite-dimensional subspaces $\V_\ga$ such that $\Delta (\V_\ga )
\subset \V_\ga $ and $\V_\ga $ is orthogonal to $\V_\gb$ for $\ga \neq \gb$.
Then $\Delta$ can be diagonalized
and it is easy to see that
$\ker\, \Omega / \Delta ({\ker \,\Omega})$ = $\ker \Delta  \cap
\ker \Omega$. Therefore, for this case,
\be
H(\Omega )\subset
\ker \,\Delta \cap  \ker\,\Omega\,.
\ee
This formula is useful in the practical analysis because
it is usually  simpler to compute $\ker \,\Delta$ than to
find  the cohomology $H(\Omega )$ directly.

Let us denote $dX^{\gm \gn }$  by $\kk^{\gm \gn }$. Then
  $\kk^{\gm \gn }=\kk^{\gn \gm }$,
$\kk^{\gm \gn}\kk^{\ga \gga }=-\kk^{\ga \gga}\kk^{\gm \gn }$
and
$$
\f{d}{d \kk^{\ga \gga}} \kk^{\gm \gn}
=\half \left(\gd^{\gm}_{\ga} \gd^{\gn}_{\gga}
+\gd^{\gn}_{\ga}\gd^{\gm}_{\gga}\right) -
\kk^{\gm \gn }\f{d}{d \kk^{\ga \gga}}.
$$
Let $\V_1$ and $\V_2$ be the linear spaces spanned by
various  polynomials $P(z,\xi)$ and  $P(z,\bar{z},\xi)$, respectively.
Let the rank 1 and rank 2 operators $\gs_-$ acting in $\V_1$
and $\V_2$ be denoted $\Omega_{1}$ and $\Omega_{2}$, respectively.
(For the future convenience  we have changed notations,
replacing  $y^\ga$ with $z^\ga$.) Defining
 the conjugated operators $\Omega^*_{1,2}$ with respect
 to the natural Fock type scalar products in $\V_{1,2}$ we have
\bee\nn
\Omega_1=
 \kk^{\ga \gga}
\f{\p}{\p z^{\ga}}
\f{\p}{\p z^{\gga} }\,,\qquad
\Omega^*_1=
z^{\ga}z^{\gga}
\f{d}{d \kk^{\ga \gga}}\,,
\eee
and
\bee
\nn
%here
\Omega_2=2
 \kk^{\ga \gga}
\f{\p}{\p z^{\ga}}
\f{\p}{\p \bz^{\gga} }\,,\qquad
%here
\Omega^*_2=2
z^{\ga}\bz^{\gga}
\f{d}{d \kk^{\ga \gga}}.
\eee
One gets
\bee \label{Delta1}
\Delta_1=
2\kk^{\gm \gn }\f{d}{\p \kk^{\gm \gn}}+
4\kk^{\gm \ga }
\f{d}{d \kk^{\gm \gb}}
z^{\gb}\f{\p}{\p z^{\ga}}+
z^{\gb}z^{\ga}
\f{\p}{\p z^{\ga}}\f{\p}{\p z^{\gb}}\,,
\eee
\bee \label{Delta}
\Delta_2=
4\kk^{\gm \gn }\f{d}{\p \kk^{\gm \gn}}+
4\kk^{\gm \ga }
\f{d}{d \kk^{\gm \gb}}
(z^{\gb}\f{\p}{\p z^{\ga}}+
\bz^{\gb}\f{\p}{\p \bz^{\ga}}) \\ \nn
+ 2z^{\ga}\f{\p}{\p z^{\ga}}\bz^{\gb}\f{\p}{\p \bz^{\gb}}+
2z^{\gb}\bz^{\ga}
\f{\p}{\p z^{\ga}}\f{\p}{\p \bz^{\gb}}\,.
\eee

Let $\V_i =\sum_p \oplus \V_i^p$, where $\V_i^p \subset \V_i$ is the
subspace of degree $p$ homogeneous polynomials in $\xi$, i.e.
\{$P(z,\ldots, \kk ) \in \V^p_i$: $ P(z,\ldots, \mu \kk )=
\mu^p P(z,\ldots, \kk )$\}.
The cohomology groups $H^p(\gs_-)$
for rank $r$ operators $\gs_-$ belong to $\ker \Delta_r \cap \V^p_r$.

Let us introduce auxiliary spaces $W_1$ and $W_2$  of polynomials
$P(z,t )$ and $ P(z,\bz ,t )$, respectively, where $t^\ga$ is an
auxiliary commuting variable. Let $W_i =\sum_p \oplus W_i^p$,
\{$P\in W_i^p \,: P(z,\ldots, \mu t )=
\mu^{p} P(z,\ldots, t )$\}.
For the cases of $\V_i^p$ with $p= 0$ or 1 associated with the
cohomology groups $H^0_r$ and $H^1_r$, the obvious isomorphisms
$\V_i^0 = W_i^0$ and $\V_i^1 = W_i^2$ take place. In practice this allows one
to  replace  $\kk^{\ga \gb}$ by $t^\ga t^\gb$ and
$\f{d}{d \kk^{\ga \gb}}$ by $\half
\f{\p}{\p t^{\ga}}\f{\p}{\p t^{\gb}}$ (for polynomials at most linear
in $\kk$ it does not  matter that $\kk$ is anticommuting).
In these variables,
the  operators $\Delta_{1,2}$  acquire the form
\bee \label{{Ups_z1}}
%here
\Delta_1=
(t^{\gm}\f{\p}{\p t^{\gm}}-1)
(2t^\ga \f{\p}{\p z^\ga} z^\gb \f{\p}{\p t^\gb}
 -
t^{\gb}\f{\p}{\p   t^{\gb}}) +
(z^{\ga}\f{\p}{\p z^{\ga}}-1)z^{\gb}\f{\p}{\p z^{\gb}}
\eee
\bee  \nn \label{{Ups_z}}
%here
\half \Delta_2=
(t^{\gm}\f{\p}{\p t^{\gm}}-1)(t^\ga \f{\p}{\p z^\ga} z^\gb \f{\p}{\p t^\gb} +
t^\ga \f{\p}{\p \bz^\ga} \bz^\gb \f{\p}{\p t^\gb}
- t^{\gb}\f{\p}{\p   t^{\gb}})  +\\
+\bz^\ga \f{\p}{\p z^\ga} z^\gb \f{\p}{\p \bz^\gb}
+ (z^{\ga}\f{\p}{\p z^{\ga}}-1)\bz^{\gb}\f{\p}{\p \bz^{\gb}}
\,.
\eee

\subsection*{Rank 1}

The operators
\bee
\label{cart2}
e_1=t^\gga \frac{\ptl}{\ptl {z}^\gga}\,,\quad
f_1=z^\gga \frac{\ptl}{\ptl {t}^\gga}\,,\quad
h_1=
t^\gga \frac{\ptl}{\ptl {t}^\gga}
-z^\gga \frac{\ptl}{\ptl {z}^\gga}\,
\eee
form  the Lie algebra $sl_2$. The operator
\be
\label{cent}
n_1=t^\gga \frac{\ptl}{\ptl {t}^\gga}
+z^\gga \frac{\ptl}{\ptl {z}^\gga}
\ee
 extends it to  $gl_2$
 with the carrier space $W_1$.
Since the action of $gl_2$ in $W_1$ leaves a  degree of a polynomial
invariant, $W_1$ decomposes into the infinite
direct sum of finite-dimensional $gl_2-$submodules.
Irreducible finite-dimensional submodules are characterized by their
$sl_2$ lowest (equivalently, highest) weights and eigenvalues of
the central element of $gl_2$ (\ref{cent}). The weight basis  of
$sl_2$ is convenient because the operator $\Delta_1$ turns
out to be diagonal in this basis.

Indeed, let $p \in W_1$ be some lowest  vector
with fixed weight and eigenvalue of
the central element of $gl_2$, i.e.
\be
f_1 p =0\,,\qquad h_1 p = \gm p\, \qquad n_1 p = \nu p\,.
\ee
 {}From (\ref{cart2}), (\ref{cent}) it follows  that $p$ has a form
\bee
p_{(q,r)}=
c_{\ga_1\dots \ga_q, \gb_1 \dots \gb_r} z^{\ga_1}\dots  z^{\ga_q}  t^{\gb_1}
\dots  t^{\gb_r}\,,\qquad \gm = r-q\,,\quad \nu =r+q\,,
\eee
where $c_{\ga_1\dots \ga_q, \gb_1 \dots \gb_r}$ is some
tensor with the symmetry properties of the  Young tableau
\begin{picture}(43,15)%(0,05)
\put( 9,12){\scriptsize   q}    % melkaja, two rows
\put( 9,-5){\scriptsize   r}    % melkaja, two rows
{\linethickness{.250mm}
\put(00,10){\line(1,0){30}}
\put(00,05){\line(1,0){30}}
\put(00,00){\line(1,0){15}}
\put(00,00){\line(0,1){10}}%
\put(05,00.0){\line(0,1){10}}
\put(10,00.0){\line(0,1){10}}
\put(15,00.0){\line(0,1){10}}
\put(20,05.0){\line(0,1){05}}
\put(25,05.0){\line(0,1){05}}
\put(30,05.0){\line(0,1){05}}
}
\end{picture}.
%\vskip.1cm
As a result, every polynomial $P(z,t)$ is a linear combination of
the polynomials
\bee    \nn
\label{psl2}
P_{(q\,r)}^{(b)} (z,t)
=\left(e_1
%t^\gga \frac{\ptl}{\ptl {z}^\gga}
\right)^b
p_{(q\,r)}(z,t)
\eee
with various $b,q$ and $r$.
Using  that $[ f_1 , (e_1)^b]=(e_1)^{b-1}b(-h_1-b+1)$, one finds
\bee
\Delta_1\, P_{(q\,r)}^{(b)} =\left( 2 e_1 f_1  - (n_1-1) h_1 \right)\,
(e_1)^b p_{(q\,r)}
%\\ \nn
= \lambda_{1}\, P_{(q\,r)}^{(b)}\,,
\eee where
\be
\label{lam1}
\lambda_{1}=(2(q-b-r+1)b+(q-b)(q-b-1)-(b+r)(b+r-1))\,.
\ee
Note that, because the representations under
consideration are finite-dimensional, the polynomials
$P_{(q\,r)}^{(b)}$  with $b> q-r$ vanish, i.e. the inequality
\bee
\label{con03}  q-b-r&\ge& 0
\eee
is true.
The degree of $P_{(q\,r)}^{(b)}(z,t)$
in $t$ is equal to $r+b$. As a result we have $r+b=0$ and
 $r+b=2$ in the subspaces $W_1^0$ and  $W_1^2$ associated with
rank 1 cohomology groups $H^0 (\gs_- )$ and $H^1 (\gs_- )$, respectively.

%\subsubsection*{Rank 1  group $H^0 $}
\subsubsection*{${\bf H^0 }$}
{Consider plynomials $P_{(q\,r)}^{(b)} $ associated with rank $1$
$0$-forms.
In this case $b=r=0$. From (\ref{lam1})
and the equation  $\lambda_1=0$ it follows that $q(q-1)=0$.
So, the rank 1 cohomology group $H^0(\Omega_1 )$ is spanned by }
$$
c\,\,, {\quad } c_{\ga}z^{\ga}.$$

\subsubsection*{${\bf H^1 }$}
%\subsection*{rank $1$ $H^1$}
Consider polynomials $P_{(q\,r)}^{(b)} $ corresponding to the rank 1
1-forms.
Then $b+r=2$. Using (\ref{lam1}) one observes that
the equation $\lambda_1=0$ amounts to
$$2(q-1)b+(q-b)(q-b-1)-2=0.$$
{}From (\ref{con03}) it follows that
$q\ge 2$ .  \\
I.\,  Let $b > 0$. Then $2(q-1)b\ge2$. As a result the only allowed
solution is $q=2, b=1$.\\
II. Let $b=0$.  The only allowed
solution is $q=2$.

It is easy to see
that  $P_{(2,1)}^{(1)}$ and $P_{(2,2)}^{(0)}$
%of $\Delta_1 f=0$
are $\Omega_1$ closed.
Since exact forms are symmetric in all variables,
the obtained solutions belong to the nontrivial cohomology
classes. Thus the rank 1 cohomology group $ H^1
(\Omega_1 )$ is spanned by
$$
c_{\ga \gb_1,\gb_2}\,\,z^\ga  \kk^{\gb_1  \gb_2}, \qquad
c_{\ga_1 \ga_2,\gb_1  \gb_2}\,\,z^{\ga_1} z^{\ga_2} \kk^{\gb_1  \gb_2}.   $$

\subsection*{Rank 2}

The rank 2 case will be analyzed in terms of
 $gl_3$ generated by the operators
\bee \label{cart3}
e_1=t^\gga \frac{\ptl}{\ptl {z}^\gga}\,,\quad
f_1=z^\gga \frac{\ptl}{\ptl {t}^\gga}\,,\quad
e_2=z^\gga \frac{\ptl}{\ptl {\bz}^\gga}\,,\quad
f_2=\bz^\gga \frac{\ptl}{\ptl {z}^\gga}\,,\qquad\qquad\\ \nn
h_1=
t^\gga \frac{\ptl}{\ptl {t}^\gga}
-z^\gga \frac{\ptl}{\ptl {z}^\gga}\,,\quad
h_2=
z^\gga \frac{\ptl}{\ptl {z}^\gga}-
\bz^\gga \frac{\ptl}{\ptl {\bz}^\gga}\,,\quad
n_2=t^\gga \frac{\ptl}{\ptl {t}^\gga}
+z^\gga \frac{\ptl}{\ptl {z}^\gga}
+\bz^\gga \frac{\ptl}{\ptl {\bz}^\gga}.
\eee
Again,  $W_2$ decomposes into the infinite direct sum
of irreducible finite-dimensional $gl_3-$modules.
Let $p \in W_2$ be some lowest  vector
with fixed weight and eigenvalue of
$n_2$, i.e.
\bee
f_1 p =0\,,\qquad f_2 p =0\,,\qquad
h_1 p = \gm_1 p\,,\qquad
h_2 p = \gm_2 p\, \qquad n_2 p = \nu p\,.
\eee
 {}From (\ref{cart3}) it follows  that $p$ has a form
\bee
p_{(s,q,r)}=
c_{\gga_1\dots \gga_s, \ga_1\dots \ga_q, \gb_1 \dots \gb_r}
\bz^{\gga_1}\dots \bz^{\gga_s}
z^{\ga_1}\dots  z^{\ga_q}  t^{\gb_1}
\dots  t^{\gb_r}\,\\ \nn \gm_1 = q-s\,,\qquad \gm_2 = r-q\,,
\qquad \nu =r+q+s\,,
\eee
where $c_{\gga_1\dots \gga_s, \ga_1\dots \ga_q, \gb_1 \dots \gb_r} $
is some tensor with the symmetry properties of the Young tableau
\begin{picture}(48,20)(0,05)
\put( 9,17){\scriptsize   s}    % melkaja, three rows
\put(43, 5){\scriptsize   q}    %
\put( 9,-5){\scriptsize   r}    %
{\linethickness{.250mm}
\put(00,15){\line(1,0){40}}
\put(00,10){\line(1,0){40}}
\put(00,05){\line(1,0){30}}
\put(00,00){\line(1,0){15}}
\put(00,00){\line(0,1){15}}%
\put(05,00.0){\line(0,1){15}}
\put(10,00.0){\line(0,1){15}}
\put(15,00.0){\line(0,1){15}}
\put(20,05.0){\line(0,1){10}}
\put(25,05.0){\line(0,1){10}}
\put(30,05.0){\line(0,1){10}}
\put(35,10.0){\line(0,1){05}}
\put(40,10.0){\line(0,1){05}}
\put(40,10.0){\line(0,1){05}}
}
\end{picture} .
Every polynomial $P(z,\bz,t)$ is some linear combination of the \\ \\
polynomials of the form
\bee\nn
%\label{psl3}
P_{(s\,q\,r)}^{(b\,a\,k)}
=\left(e_1
%t^\gga \frac{\ptl}{\ptl {z}^\gga}
\right)^b
\left([e_1, e_2]
%t^\gga \frac{\ptl}{\ptl {\bz}^\gga}
\right)^a
\left(e_2
%z^\gga \frac{\ptl}{\ptl {\bz}^\gga}
\right)^k
p_{(s\,q\,r)}
\eee
with various $b,\,a,\,k$. An elementary computation shows that
\bee \label{Del_spek}
%here
\half \Delta_2\, P_{(s\,q\,r)}^{(b\,a\,k)}=
\lambda_2\,
P_{(s\,q\,r)}^{(b\,a\,k)},\eee where
\bee
\label{Lam}\nn   \lambda_2=
(q+k-a-r)b-b(b-1)+(s-k-r)a +k(s-k-q+1)
\\
+ a(2-a)+(s-k-a-1)(q+k-b)-(a+b+r)(a+b+r-1).
\eee
Because all representations of $sl_3$ under
consideration are finite-dimensional
the following inequalities take place
\bee
\label{con2}  s-k-a&\ge& 0\,,\\
\label{con3}  q+k-b&\ge& 0\,,
\eee
\be
\label{con4}  q \ge r\,,
\ee
\be
\label{con1}  s-k-q\ge 0\,.
\ee
Eqs. (\ref{con2}), (\ref{con3}) manifest the simple fact that
the degrees of $z$ and $\bz$ are nonnegative. Eq. (\ref{con4}) is obvious
from the Young tableau representation of the lowest vector of $sl_3$.
Eq.(\ref{con1}) is understood  analogously taking into account
the explicit form of $e_2$  in (\ref{cart3}).

\subsubsection*{${\bf H^0 }$}
%\subsection*{rank $2$ $H^0$}
Consider polynomials $P_{(s\,q\,r)}^{(b\,a\,k)}$
corresponding to the rank 2 0-forms.
Using (\ref{Lam}) along with the fact that
 $a=b=r=0$ for the case of 0-forms
 one finds that the equation $\lambda_2=0$ amounts to
\bee \label{Ups0}
k(s-k-q+1)+ (s-k-1)(q+k)=0.
\eee
{}From (\ref{con1}) one has
$(s-q-k+1)>0$ or, equivalently,  $s-k-1\ge q-1$.\\
I.\, Let  $s-k=0$. Then $-kq-q=0$ and, therefore,  $q=0$. \\
II. Let $(s-k-1)\ge 0$. Then (\ref{Ups0}) requires
$(k+q)(s-k-1)=0$  and $k(s-q-k+1)=0$. Therefore, $k=0$
and  $q(s-1)=0$.\\
As a result (\ref{Ups0}) has  the following solutions:
  $\quad \{s=k\geq 0, q=0\};
 \quad \{k=q=0, s\geq 0\}; \quad  \{k=0, q=s=1\}$.
One can check that
 $P_{(s,0,0)}^{(0,0,s)}$, $P_{(s,0,0)}^{(0,0,0)}$ and $P_{(1,1,0)}^{(0,0,0))}$
are    $\Omega_2$-closed.
Thus the cohomology group $ H^0 (\Omega_2 )$
is parametrized by the following $0-$forms:
$$
c,         \qquad
c_{\ga, \gb}\,z^{\ga}\bar{z}^{\gb} ,\qquad
{c}_{\ga_1 \ldots \ga_n}\,
z^{\ga_{1}}\dots z^{\ga_{n}}      , \qquad
\bar{c}_{\ga_1 \ldots \ga_n}\,
\bz^{\ga_{1}}\dots \bz^{\ga_{n}},       \qquad n>0 .
$$

\subsubsection*{${\bf H^1 }$}
%\subsection*{rank $2$ $H^1$}
Consider polynomials
 $P_{(s\,q\,r)}^{(b\,a\,k)} $   corresponding to rank 2  1-forms.
In this case   $a+b+r=2$.  Then, taking into account  (\ref{Lam}),
the  equation $\lambda_2=0$ amounts to
 \be \label{Ups}
(q+k)(b-1)+(s-k+b)a+k(s-k-q+1)+(s-k-a)(q+k-b)=2 .
\ee
The problem is to solve (\ref{Ups})
in nonnegative integers at the conditions  (\ref{con2})--(\ref{con1})
along with the two obvious conditions
\bee
\label{con5}  a+b& \le& 2,\\
\label{con6} s& \ge& 2. %ne mozhet byt' stolbcom-stepeni 2 po t
\eee
(The meaning of (\ref{con6}) is that,  to have polynomials
bilinear in $t^\ga$ used to describe 1-forms, the first row of the
respective Young tableaux must contain at least two cells.)

{}From (\ref{con5}) it follows  that there are six  cases:
$\{a=0, b=0\}$,\,
$\{a=1, b=1\}$,\,
$\{a=0, b=1\}$,\,
$\{a=0, b=2\}$,\,
$\{a=1, b=0\}$ and
$\{a=2, b=0\}$. Note, that only four of these six cases are
essentially different because the interchange $z \leftrightarrow \bz$
is equivalent to $a\leftrightarrow b$. \\
I.\, Let $a=b=0$ . Then (\ref{Ups}) has the form
\bee \label{Ups00}
k(s-k-q+1)+(s-k-1)(q+k)&=&2\,.
\eee
{}From (\ref{con1}), (\ref{con4}) it follows  that
$s-k\ge q\ge 2$ and, therefore, (\ref{Ups00}) requires
 $q=2$, $s=2$, $k=0$ .\\
II.\, Let  $b=1$,     $a=1$. Then (\ref{Ups}) has the form
\bee \label{Ups11}
k(s-k-q+1)+(s-k-1)(q+k)&=&0. \eee
{}From (\ref{con2}) and (\ref{con1}) one can see that
$k(s-k-q+1) \geq 0$ and $(s-k-1)(q+k)\geq 0$. Therefore,
(\ref{Ups11}) requires $k(s-k-q+1) = 0$ and $(s-k-1)(q+k)= 0$.
The first condition requires
$k=0$. Then from (\ref{con3}), (\ref{con6}) one finds that the second
one has no admissible solutions.\\
%III. Let $b\ne0$ .  Then all terms in the left hand side of (\ref{Ups}) are
%nonnegative by virtue of (\ref{con1})---(\ref{con5}).\\
III. Let $b=1$, $a=0$. Then (\ref{Ups}) has the form
\bee \label{Ups10}
k(s-k-q+1)+(s-k)(q+k-1)&=&2\,.
\eee
Since $a+b+r=2$ we have   $r=1$. {}From (\ref{con4}) it follows that
$ q\ge1$. {}From (\ref{con1}) we get $ s-k\ge q\ge 1$.
As a result, the two terms on the left hand side of (\ref{Ups10}) are
non-negative. So, there are three different cases:\\
A. $k(s-k-q+1)=0$ ,
$(s-k)(q+k-1)=2$. The only admissible solution is $k=0,$  $s=2,$ $q=2.$\\
B.  $k(s-k-q+1)=1$, $(s-k)(q+k-1)=1$. The only solution is
$k=1$, $s=2$, $q=1$.\\
C.  $k(s-k-q+1)=2$, $(s-k)(q+k-1)=0$. It is easy to see that
these equations have no admissible solutions.\\
IV.  Let $b=2$, $a=0$.  Then (\ref{Ups}) has the form
\bee \label{Ups20}
(q+k) +k(s-k-q+1)+(s-k)(q+k-2)&=&2.
\eee
Using (\ref{con3}) one gets $ q+k=2$,
$k(s-k-q+1)=0$. Using (\ref{con1}) one gets
$k=0$, $q=2$, $\forall s$.

Instead of considering the rest two cases with
$\{a=1, b=0\}$ and
$\{a=2, b=0\}$ we simply add the
conjugated forms resulting from the interchange $z \leftrightarrow
\bz$ to  get the full list
of $\Delta_2$ zero modes.
It is easy to see that the obtained solutions
are    $\Omega_2$-closed but not exact.
As a result, the cohomology group $H^1 (\Omega_2 )$
is spanned by
$$
c_{\gga_1 \gga_2,\,\ga,\,\gb}\,\,
z^{\ga}  \bar{z}^{\gb} \kk^{\gga_1\gga_2}, \qquad
c_{\gga_1\gga_2,\,\gb_1 \gb_2,\,\ga}\,\,
            z^{\ga}\bz^{\gb_1}\bz^{\gb_2}\kk^{\gga_1\gga_2},\qquad
$$ $$
c_{\gga_1\gga_2,\,\ga_1 \ga_2,\,\gb_1 \gb_2}\,\,
z^{\ga_1}z^{\ga_2}\bz^{\gb_1}\bz^{\gb_2}\kk^{\gga_1\gga_2}, \quad
c_{\gga_1\gga_2,\,\ga_1 \ga_2,\,\gb}\,\,
z^{\ga_1}z^{\ga_2}\bz^{\gb}\kk^{\gga_1\gga_2},
$$ $$
{c}_{\ga_1 \dots \ga_n,\gga_1\gga_2}\,\,
{z^{\ga_1}\dots z^{\ga_n} \kk^{\gga_1\gga_2}} , \quad
{\bar c}_{\ga_1 \dots \ga_n,\gga_1\gga_2}\,\,
{\bz^{\ga_1}\dots \bz^{\ga_n}} \kk^{\gga_1\gga_2}\,,\quad  n\ge 2\,.
$$

\end{document}